\documentclass[aps,
		pra,
		reprint,
		twocolumn,
		superscriptaddress,
		shortbibliography,
		nofootinbib,
		floatfix,
		notitlepage
		]{revtex4-2}
\pdfoutput=1
\usepackage[utf8]{inputenc}
\usepackage[english]{babel}
\usepackage[T1]{fontenc}
\usepackage{amsmath}
\usepackage{hyperref}

\usepackage{tikz}
\usepackage{lipsum}

\usepackage{bm}
\usepackage{bbm}
\usepackage{bbold}
\usepackage{slashed}
\usepackage{mathrsfs}

\usepackage{dsfont}

\usepackage{amsmath,amssymb}
\usepackage{graphicx,accents}
\usepackage{graphicx,epstopdf}
\usepackage[normalem]{ulem}

\usepackage{hyperref}

\hypersetup{colorlinks=true}

\newcommand{\trm}[1]{\textrm{#1}}
\newcommand{\ud}{\mathrm{d}}
\newcommand{\LCm}{{\scriptscriptstyle -}}
\newcommand{\LCp}{{\scriptscriptstyle +}}
\newcommand{\LCpm}{{\scriptscriptstyle \pm}}

\newcommand{\LCperp}{{\scriptscriptstyle \perp}}

\newcommand{\bra}[1]{\langle #1 |}
\newcommand{\ket}[1]{|#1\rangle}

\newcommand{\figref}[1]{Fig. \ref{#1}}
\newcommand{\figrefa}[1]{Fig. \ref{#1}(a)}

\newcommand{\eqnref}[1]{Eq. (\ref{#1})}
\newcommand{\sxnref}[1]{Sec. \ref{#1}}

\newcommand{\sscript}{\scriptscriptstyle}
\newcommand{\bi}{\begin{itemize}}
\newcommand{\ei}{\end{itemize}}
\newcommand{\xiE}{\mathcal{E}}

\newcommand{\vsigma}{\varsigma}

\newcommand{\vtheta}{\vartheta}
\newcommand{\vphi}{\varphi}
\newcommand{\mC}{\mathcal{C}}
\newcommand{\maf}[1]{\mathfrak{#1}}

\begin{document}

\title{Entanglement and pair production in intense electromagnetic fields}

\author{S. Tang}
\affiliation{College of Physics and Optoelectronic Engineering, Ocean University of China, Qingdao, Shandong, 266100, China}
\author{B. M. Dillon}
\affiliation{ISRC, Ulster University, Derry, BT48 7JL, Northern Ireland}
\author{B. King}
\affiliation{Centre for Mathematical Sciences, University of Plymouth, Plymouth, PL4 8AA, United
Kingdom}
\email{b.king@plymouth.ac.uk}

\begin{abstract}
We investigate the spin correlations between electron-positron pairs created from a photon when it scatters in a high-intensity laser pulse via the nonlinear Breit-Wheeler process. We find that the spin states of the generated electron-positron pair can exhibit strong entanglement, with the degree being sensitive to the photon energy, laser intensity, and the relative polarisation of the photon and laser pulse. Photons with a high degree of polarisation can create strongly entangled pairs, and this entanglement can be maintained in the high-intensity (non-perturbative) regime. We find that if the photons are provided by a Compton source, strongly spin-entangled electron-positron pairs can be generated
with technology available today.
\end{abstract}

\maketitle

\section{Introduction}

Entanglement is one of the most interesting phenomena predicted by quantum physics, and central to quantum information and quantum computing~\cite{NielsenChuang2010}.
Many experiments have measured violations of the Bell inequality \cite{bell_einstein-podolsky-rosen_1964}, including the 1972 \cite{PhysRevLett.28.938} and 1982 \cite{PhysRevLett.49.1804,PhysRevLett4991} experiments using entangled photons that received the 2022 Nobel prize, and others since \cite{PhysRevLett4991,PhysRevLett791,PRL99131802,science1130886,pfaff2013demonstration,lee2011entangling}.
There has been significant interest recently in studying entanglement in top-pairs at the Large Hadron Collider (LHC) \cite{afik2021entanglement,PRLett161801,Aoude:2022imd,Afik:2022kwm,Dong:2023xiw,Han:2023fci}, providing insights into quantum effects at high-energies and potential probes of new physics \cite{Fabbrichesi:2025ywl}.
ATLAS and CMS have published observations of entanglement in the leptonic final-states of $t\bar{t}$ production \cite{atlas2024observation,collaboration_observation_2024-1,CMS:2024zkc}, but there are many more systems where this effect can be measured \cite{Fabbrichesi:2023cev,Fabbrichesi:2022ovb,Aguilar-Saavedra:2022wam,Ashby-Pickering:2022umy,Sakurai:2023nsc,Aguilar-Saavedra:2024whi} (see \cite{Barr:2024djo,afik2025quantuminformationmeetshighenergy} for a review).
The high-intensity/strong-field regime is yet another, less studied, scenario where we can probe quantum entanglement.

The production of an electron-positron pair as a high-energy photon collides with an intense laser pulse is often referred to as nonlinear Breit-Wheeler (NBW) pair production~\cite{breit34,Reiss1962,nikishov64,Fedotov:2022ely,RMP2022_045001}.
The process has been indirectly measured as a subprocess of nonlinear trident \cite{burke97,Nielsen:2022bws} and theoretically studied in various types of laser field~\cite{TangPRA2021,PRA2012_052104,PRL2012_240406,PRA2013_062110,PRD013010,PRD2016_053011,PRA2013_052125,Titov:2015tdz,EPJD2020Titov,PRA2014052108,Titov:2018bgy,ilderton2019coherent,Ilderton:2019vot,king2020uniform,PRD076017}. Direct measurement of NBW is an identified goal of modern-day laser-particle experiments~\cite{Abramowicz:2021zja,E320_2021,chen22,LUXE:2023crk}. In such experiments, the intensity parameter, $\xi$, quantifies the effective coupling between the laser field and the pair.
When the intensity is low ($\xi\ll 1$), the interaction is \emph{perturbative} in $\xi$ and well-approximated by including just the channel involving the minimum number of interactions with the laser background to reach the energy-momentum threshold for pair-creation. Since the number of interactions is typically much larger than unity, the perturbative limit is highly nonlinear in $\xi$. (At very low values of $\xi$, pulse envelope effects can in principle change this interpretation; these are interesting products of theory and we include them in our analysis but this region is practically inaccessible to experiment.)

As the intensity is increased, so too is the contribution of channels that involve more laser photons.
Values of $\xi \sim O(1)$ can routinely be produced in the lab; at such intensities all orders of interaction between the laser and the pair must be included in calculation, which is sometimes referred to as \emph{non-perturbativity at small coupling}.

The motivation for this paper is to investigate the effect that this non-perturbative interaction has on the degree of entanglement in states generated in laser-particle collisions. The process of NBW pair creation is used as a canonical example to study the concurrence of the spin-polarisation in the generated electron-positron pair. Several works have investigated entanglement in laser-particle interactions, most notably in the process of Compton scattering \cite{Schutzhold:2008zza,PhysRevA.80.053419,Lotstedt:2013uya,Zhang:2022qaa,deVos:2023pen}.
Entanglement has been less studied in pair-creation although it was investigated in the Breit-Wheeler process in \cite{FEDOROV2006413} for the case of colliding two finite wavepackets of photons with frequencies exceeding $mc^{2}$, i.e. the rest mass of an electron or positron.
In another more recent paper entanglement in Breit-Wheeler production was studied as a component of the trident process where the intermediate photon is on-shell \cite{Roshchupkin:2025pej}, with the main focus being on the general kinematics of the final-state particles.
The challenge of measuring the polarisation of free-travelling electron-positron pairs to determine their entanglement has been studied recently in \cite{Gao:2025kdi}, where they propose using secondary scatterings to determine the polarisation of each electron/positron for pairs with energies between $1$ and $10$ GeV.

Our work in this paper is the first study of entanglement in strong electromagnetic backgrounds to NBW pair creation in an intensity range spanning the perturbative $\xi \ll 1$ to non-perturbative $\xi\gg 1$, in an energy range from the nonlinear $\eta \ll 1$ to the linear $\eta > 2(1+\xi^{2})$ (where $\eta$ is the photon energy parameter), considering a range of photon polarisations creating pairs in a  circularly-polarised and linearly-polarised background.
We choose the concurrence observable to measure the strength of entanglement, which we derive from the density matrices of the final state.
Thereby, we derive and present local approximations (locally constant field (LCFA) and locally monochromatic (LMA)) of the density matrix that can be employed in numerical simulations and benchmark these approximations against direct evaluation of full QED result in a plane wave pulse background. We consider mono-energetic photons as well as a two-stage scenario where photons are produced via nonlinear Compton scattering before colliding with a second laser pulse to create pairs.

The paper is organised as follows. In Sec. 2, the theoretical model is presented and information on the local approximation is given. In Sec. 3 the numerical results for the mono-energetic (Sec 3.1) and the two-stage set-up (Sec 3.2) are presented and discussed. The paper is concluded in Sec. 4. Appendix A gives technical details of the spin quantisation used, and Appendix B gives further details about calculation of the spectrum from the Compton photon source.

\section{Theoretical model}~\label{Sec_2}
\begin{figure}[h!!]
\center{\includegraphics[width=6cm]{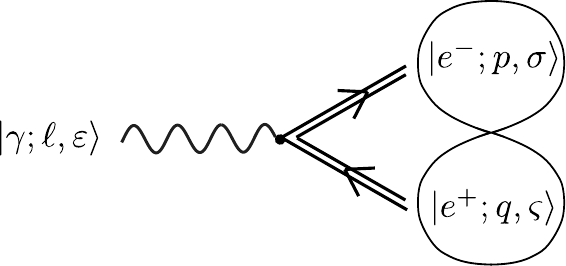}}
\caption{The nonlinear Breit-Wheeler process with momentum and polarisation labels of each particle. The double solid lines represent fermionic Volkov states and the interlinked circles denote the emitted particles are entangled.}
\label{fig:diag1}
\end{figure}

In the NBW process, a photon in the initial state is converted to an electron and a positron in the final state. We represent this with state vectors:
\[
|\trm{in}\rangle = |\gamma;\ell,\varepsilon\rangle; \quad |\trm{out}\rangle = \Big|\left[e^{\LCm};p,\sigma\right],\left[e^{\LCp};q,\varsigma\right]\Big\rangle,
\]
where a photon with momentum $\ell$ and polarisation $\varepsilon$ is converted to a spin-entangled two-particle state comprising an electron with momentum $p$ and spin state $\sigma$ and a positron with momentum $q$ and spin state $\varsigma$ as depicted in Fig.~\ref{fig:diag1}.
The in and out states are related via the scattering operator, $\hat{S}$:
\[|\trm{out} \rangle = \hat{S}\, |\trm{in} \rangle.\]
To investigate the spin entanglement between the electron and positron, we formulate the spin correlation density matrix, $\rho_{b} = \ket{\trm{out}}\bra{\trm{out}}$ by tracing out their momentum states:
\begin{align}
\rho_{b}=&\frac{1}{\trm{P}_{b}}\int\frac{V\ud^{3}p}{(2\pi)^3}\frac{V\ud^{3}q}{(2\pi)^3} \begin{bmatrix}
\langle e^{\LCm};p,+| \\
\langle e^{\LCm};p,-|
\end{bmatrix}\otimes
\begin{bmatrix}
\langle e^{\LCp};q,+| \\
\langle e^{\LCp};q,-|
\end{bmatrix}\nonumber\\
&~~~~~~~~~~~~~~~\hat{S}|\gamma;\ell,\varepsilon\rangle\langle\gamma;\ell,\varepsilon|\hat{S}^{\dagger}\nonumber\\
&{\begin{bmatrix}
|e^{\LCm};p,+\rangle & |e^{\LCm};p,-\rangle
\end{bmatrix}\otimes}\nonumber\\
&\hspace{1cm} {\begin{bmatrix}
|e^{\LCp};q,+\rangle & |e^{\LCp};q,-\rangle
\end{bmatrix}}
\label{Eq_finaldensmatrix}
\end{align}
normalised with the total probability $\trm{P}_{b}$
\begin{align}
\trm{P}_{b}=&\sum_{\sigma,\varsigma=\pm} \int\frac{V\ud^{3}p}{(2\pi)^3}\frac{V\ud^{3}q}{(2\pi)^3}  \nonumber \\
& \left|\,\left(\langle e^{\LCm};p,\sigma|\otimes\langle e^{\LCp};q,\varsigma|\right)~\hat{S}~|\gamma;\ell,\varepsilon\rangle\,\right|^{2},
\label{Eq_prob0}
\end{align}
so that $\trm{Tr}(\rho_{b})=1$.
The $4\times 4$ density matrix $\rho_{b}$ contains the full information of the paired particles' spin states.
By partially tracing over the spin freedom of the positron (electron), the spin density matrix of electron (positron) can be acquired as $\rho^{\LCm} = \trm{Tr}_{\vsigma}(\rho_{b,\sigma,\vsigma;\sigma',\vsigma})$ ($\rho^{\LCp} = \trm{Tr}_{\sigma}(\rho_{b,\sigma,\vsigma;\sigma,\vsigma'})$)~\cite{Tang:2022a}.
It is a principle feature of entanglement that the full denstiy matrix $\rho_{b}$ cannot in general be factorised into a direct product $\rho^{\LCp}\otimes \rho^{\LCm}$ of density matrices of each independent particle.
Furthermore, we emphasise that the selection of spin quantum axis is arbitrary and whilst it can change the expression for the components of the density matrix, it cannot affect the prediction of any physical results (see the discussion in Appendix.~\ref{App_spinbasis}).

The degree of the spin entanglement can be assessed by calculating an entanglement witness; in this work we use the concurrence, $\mC$, for this purpose. The concurrence can be written \cite{PRL802245,Barr:2024djo}:
\begin{align}
\mC= \max(0,\sqrt{e_{1}}-\sqrt{e_{2}}-\sqrt{e_{3}}-\sqrt{e_{4}})
\label{Eq_concurrence}
\end{align}
where $e_{j}$ for $j\in\{1,2,3,4\}$ are the eigenvalues, ranked in order of decreasing magnitude, of the auxiliary matrix
\begin{align}
R= \rho_{b}~(\sigma_{2}\otimes\sigma_{2})~\rho^{*}_{b}~(\sigma_{2}\otimes\sigma_{2})\,,
\label{Eq_concurrence_mat}
\end{align}
with $\rho^{*}_{b}$ a matrix having entries equal to the complex conjugate of the corresponding entries in $\rho_{b}$ and $\sigma_{2}$ is the second Pauli matrix.
The concurrence satisfies $0\leq \mC \leq 1$: if $\mC=1$, the quantum state is maximally entangled; if $\mC=0$ the state is not entangled and the density matrix can be factorised.

\subsection{Polarised NBW in pulsed plane waves}\label{Sec_Single_photon_pair_creation}
The NBW process forms a scientific aim of upcoming experiments that collide high energy photons with a high power laser pulse such as the LUXE experiment \cite{Abramowicz:2021zja}.
Because the typical energies of photons available in such set-ups are $\sim O(\trm{GeV})$ and the frequency of laser photons $\sim O(\trm{eV})$, the centre-of-mass energy of a collision $\sim O(10\,\trm{keV})$, much less than the pair creation threshold $O(\trm{MeV})$.
Therefore the NBW process can only proceed by the net absorption of high numbers photons from the laser pulse. This means calculations must typically include all orders of interaction between the laser and the produced pair. The standard approach is to calculate matrix elements in the Furry picture by solving the Dirac equation in a classical plane wave electromagnetic background to acquire fermionic Volkov states \cite{wolkow1935klasse}. The classical plane-wave background is used to represent the microscopic interaction with the laser pulse; resulting probabilities can then be integrated over the macroscopic details of the laser beam. The total probability can be calculated using an expansion of interactions between the quantised electromagnetic field and the Volkov states. Laser-particle experiments aim to measure the leading-order, tree-level contribution to the NBW process, and this is the contribution we calculate in this work. The derivation of this leading-order  process is well-documented in the literature; for the fully-polarised case, see \cite{Tang:2022a}. (For reviews of strong-field QED see \cite{ritus85,Kaminski:2009wwd,DiPiazza:2011tq,RMP2022_045001,Fedotov:2022ely,Sarri:2025qng}).

In this section, we define variables and give expressions that will be useful when discussing the local approximations and numerical results.
We begin by defining the (plane-wave) laser pulse with scaled vector potential $a^{\mu}(\phi)=|e|A^{\mu}(\phi)=(0,a_{x}(\phi),a_{y}(\phi),0)$, wavevector $k^\mu = \omega (1,0,0,-1)$ and laser phase $\phi=k\cdot x$, where $|e|$ is the charge of the positron, and $\omega$ is the laser frequency.
The photon-laser collision is characterised by the (photon) energy parameter $\eta =k\cdot \ell/m^{2}$ and the intensity parameter $\xi$, which enters in the definition of the vector potential of the laser field:
\begin{align}
a^{\mu}(\phi)=m \xi_{0}~f(\phi)~(0,\cos\phi, \mathfrak{c} \sin\phi,0)\,,
\label{Eq_Com__plane_cir1_slow}
\end{align}
where $f(\phi)$ is the pulse envelope, $\mathfrak{c}=0$ is for linear polarisation, and $\mathfrak{c}=\pm1$ denote circular polarisation with the field rotation corresponding to the polarisation state $(\epsilon_{1}+i\maf{c} \epsilon_{2})/\sqrt{2}$ with $\epsilon_{1}=(0,1,0,0)$, $\epsilon_{2}=(0,0,1,0)$.

The polarisation of the incoming photon can be fully described with a $2\times2$ density matrix as
\begin{align}
|\gamma;\ell,\varepsilon\rangle\langle\gamma;\ell,\varepsilon|=\sum_{\lambda,\lambda'= \pm}\rho_{\gamma,\sscript\lambda\lambda'}|\ell,\varepsilon_{\lambda}\rangle\langle\ell,\varepsilon_{\lambda'}|\,,
\end{align}
where the polarisation basis~\cite{baier76}:
\begin{align}
\varepsilon^{\mu}_{\LCpm}=&\epsilon^{\mu}_{\LCpm}-\frac{\ell\cdot \epsilon_{\LCpm}}{k\cdot \ell}k^{\mu}\,,
\label{Eq_phpolar}
\end{align}
is used, satisfying the relations $\ell\cdot \varepsilon_{\pm}=0$ and $k\cdot \varepsilon_{\pm}=0$, with $\epsilon_{\pm}=(\epsilon_{1}\pm i\epsilon_{2})/\sqrt{2}$.
The subscript `$\lambda=\pm$' marks the rotation direction of the polarisation state: $\lambda=+1$ is for right-hand polarisation and $\lambda=-1$ for left-hand polarisation, where the photon is right-hand polarised when the angular momentum is
parallel to the direction of its propagation, and left-handed when anti-parallel~\cite{landau4}. The photon's polarisation density matrix can be expressed explicitly as:
\begin{align}
\rho_{\gamma}=\frac{1}{2}\begin{bmatrix}
  1+\Gamma_{3} & \Gamma_{1}-i\Gamma_{2} \\
  \Gamma_{1}+i\Gamma_{2} & 1-\Gamma_{3}
\end{bmatrix}\,.
\end{align}
The (normalised) Stokes parameters ($\Gamma_{1}$, $\Gamma_{2}$, $\Gamma_{3}$) measure both the direction and degree of the photon polarisation;
$\Gamma_{3}$ is the degree of the circular polarisation, and $\Gamma_{1}$ ($\Gamma_{2}$) is the linear polarisation degree: $\Gamma_{1}=+1$ ($-1$) corresponds to the complete polarisation along $x$ ($y$)-direction and $\Gamma_{2}=+1$ ($-1$) corresponds to the complete polarisation along $45^{\circ}$ ($135^{\circ}$)-direction in the $x$-$y$ plane~\cite{landau4}.
The total polarisation degree of the photon beam is given as \mbox{$\Gamma=(\Gamma^{2}_{1}+\Gamma^{2}_{2}+\Gamma^{2}_{3})^{1/2}$} where $0\leq \Gamma\leq1$; $\Gamma=1$ means a completely polarised photon, and $\Gamma=0$ an unpolarised photon.
Again, the selection of polarisation basis is arbitrary, and selecting a different polarisation basis changes only the elements of $\rho_{\gamma}$, but not measurable physical quantities.

The electron and positron are produced in Volkov states with bispinors $u_{p,\sigma}$  and $v_{q,\varsigma}$ respectively (explicit expressions are presented in Appendix.~\ref{App_spinbasis}). We use the lightfront spin quantisation axis defined by:
\begin{align}
S^{\mu}_{p}&=\frac{p^{\mu}}{m}-\frac{m}{k\cdot p}k^{\mu}\,,
\label{Eq_lightfront_spin_quantization}
\end{align}
for the electron and with the replacement $p\to q$ for the positron.

The spin correlation density matrix and production probability can be parametrised using combinations of lightfront momenta. We use commonly-chosen variables for the NBW process: $s=k\cdot q/k\cdot \ell$ is the positron lightfront momentum fraction ($0\leq s \leq 1)$ and \mbox{$\bm{r}=(r_{x},r_{y})$}, and \mbox{$r_{x,y}= q_{x,y}/m-s\ell_{x,y}/m$} is the `transverse' positron momentum in the plane perpendicular to the laser propagating direction. Conservation of one component of lightfront momentum is expressed by $s+t=1$, where $t$ is the electron lightfront momentum fraction.

The photon-polarised probability, $\trm{P}_{b}$, for NBW can be written as \cite{TangPRD056003}:
\begin{align}
&\trm{P}_{b}=\frac{\alpha}{(2\pi\eta )^2}\int \frac{\ud s}{ts}\int\ud^{2} \bm{r} \iint \ud \phi_{1}\ud \phi_{2}~ e^{i\int_{\phi_{2}}^{\phi_{1}}\ud\phi\frac{\ell\cdot\pi_{q}(\phi)}{m^{2}\eta t}}\nonumber\\
&~~\left[h_s (\Delta a)^{2} + 1 - 2h_s \mathcal{R}\Gamma_{3} + n_{1}\Gamma_{1} + n_{2}\Gamma_{2}\right]
\label{Eq_prob_plane}
\end{align}
where $\alpha$ is the fine structure constant. The exponent is semi-classical; it contains the classical solution of the momentum of a positron in a plane electromagnetic wave: $\pi_{q}(\phi)=q^{\mu}-a^{\mu}(\phi)+[q\cdot a(\phi)/k\cdot q - a^2(\phi)/2k\cdot q] k^{\mu}$. The pre-exponent contains the kinematic variable $h_{s}=(s^2+t^2)/(4st)$, the laser-field dependent variable $\Delta a= [a(\phi_{1}) - a(\phi_{2})]/m$ and the mixed variables $\mathcal{R}=i[w_{x}(\phi_{1})w_{y}(\phi_{2})-w_{y}(\phi_{1})w_{x}(\phi_{2})]$, $n_{1} = w_{y}(\phi_{1}) w_{y}(\phi_{2})-w_{x}(\phi_{1}) w_{x}(\phi_{2}) $ and $n_{2} = -w_{x}(\phi_{1}) w_{y}(\phi_{2}) - w_{y}(\phi_{1}) w_{x}(\phi_{2})$ where \mbox{$\bm{w}(\phi)= \bm{r} - \bm{a}^{\LCperp}(\phi)/m$} with $\bm{a}^{\LCperp}=(a_{x},a_{y})$ that couple to different photon polarisation states. The variable $\mathcal{R}$ comes from the rotation of the background field and couples with the circular polarisation $\Gamma_{3}$ whereas $n_{1}$ and $n_{2}$ are the coupling of the laser's linear polarisation with the seed photon's linear polarisation degree $\Gamma_{1}$ and $\Gamma_{2}$ respectively. The corresponding spin-density matrix is:
\begin{align}
\rho_{{ b}}=&\frac{1}{\trm{P}_{b}}\frac{\alpha}{(2\pi\eta)^2}\int \frac{\ud s}{ts} \int\ud^{2} \bm{r} \iint \ud \phi_{1} \ud \phi_{2}\label{Eq_densmatrix_plane}\\
&\left[\rho^{0}_{q} + \Gamma_{1}\rho^{1}_{q} + \Gamma_{2}\rho^{2}_{q} + \Gamma_{3}\rho^{3}_{q} \right]~e^{i\int_{\phi_{2}}^{\phi_{1}}\ud\phi\frac{\ell\cdot\pi_{q}(\phi)}{m^{2}\eta t}}\,,\nonumber
\end{align}
where
\begin{align}
\rho^{0}_{q} =&\begin{bmatrix}
\frac{1}{4st} & -\frac{\tilde{w}^{*}_{2}}{4s} & \frac{\tilde{w}^{*}_{2}}{4t} & 0 \\
-\frac{\tilde{w}_{1}}{4s} & h_{s}\mathcal{A} + g_{s} \mathcal{R} &-\mathcal{A}/2 &-\frac{\tilde{w}^{*}_{1}}{4t} \\
 \frac{\tilde{w}_{1}}{4t} &-\mathcal{A}/2 & h_{s}\mathcal{A} - g_{s} \mathcal{R} & \frac{\tilde{w}^{*}_{1}}{4s} \\
0 & -\frac{\tilde{w}_{2}}{4t} & \frac{\tilde{w}_{2}}{4s} & \frac{1}{4st}
\end{bmatrix}
\label{Eq_unpolar_matrix}
\end{align}
denotes the unpolarised contribution and $\rho_{q}^{i=1,2,3}$ is the polarised contribution from the photon's polarisation component $\Gamma_{i}$:
\begin{subequations}
\begin{align}
\rho^{1}_{q}=&\begin{bmatrix}
 0 & \frac{\tilde{w}_{2}}{4t} &-\frac{\tilde{w}_{2}}{4s} &\frac{-1}{4st} \\
 \frac{\tilde{w}^{*}_{1}}{4t} & \frac{n_{1}}{2} & i n_{2}g_s - n_{1}h_{s} & \frac{\tilde{w}_{1}}{4s} \\
\frac{-\tilde{w}^{*}_{1}}{4s} &-n_{1}h_{s} - ig_s n_{2} & \frac{n_{1}}{2} &\frac{-\tilde{w}_{1}}{4t} \\
\frac{-1}{4st} & \frac{\tilde{w}^{*}_{2}}{4s} &-\frac{\tilde{w}^{*}_{2}}{4t} & 0
\end{bmatrix},\\
\rho^{2}_{q}=&i\begin{bmatrix}
 0 & -\frac{\tilde{w}_{2}}{4t}  & \frac{\tilde{w}_{2}}{4s} &  \frac{1}{4st}\\
 \frac{\tilde{w}^{*}_{1}}{4t}  &-i\frac{n_{2}}{2} & ih_{s} n_{2} - g_{s} n_{1}  & \frac{-\tilde{w}_{1}}{4s}\\
\frac{-\tilde{w}^{*}_{1}}{4s}  & g_{s} n_{1} + ih_{s} n_{2} &-i\frac{n_{2}}{2}  & \frac{\tilde{w}_{1}}{4t}  \\
 \frac{-1}{4st} & \frac{\tilde{w}^{*}_{2}}{4s}  & -\frac{\tilde{w}^{*}_{2}}{4t}  & 0
\end{bmatrix},\\
\rho^{3}_{q}=&\begin{bmatrix}
 \frac{1}{4st}  & -\frac{\tilde{w}^{*}_{2}}{4s} & \frac{\tilde{w}^{*}_{2}}{4t} & 0 \\
-\frac{\tilde{w}_{1}}{4s} &-g_{s} \mathcal{A} - h_{s} \mathcal{R} & \mathcal{R}/2 & \frac{\tilde{w}^{*}_{1}}{4t} \\
 \frac{\tilde{w}_{1}}{4t} & \mathcal{R}/2 & g_{s} \mathcal{A} - h_{s} \mathcal{R} & -\frac{\tilde{w}^{*}_{1}}{4s} \\
 0 & \frac{\tilde{w}_{2}}{4t} & -\frac{\tilde{w}_{2}}{4s} & -\frac{1}{4st}
\end{bmatrix}\,,
\end{align}
\label{Eq_polar_matrix}
\end{subequations}
$\!\!$and $\tilde{w}_{1,2} =w_{x}(\phi_{1,2}) + iw_{y}(\phi_{1,2}) $, $g_s=(s-t)/(4st)$, $\mathcal{A} = (\Delta a)^{2}/2 - 1$.

With the derived density matrix $\rho_{b}$, one can simply calculated the concurrence $\mC$ with the auxiliary matrix~(\ref{Eq_concurrence_mat}) and reveal the degree of the spin entanglement between the created electron-positron pair.
The energy (lightfront momentum fraction) and transverse-momentum dependence of the final density matrix $\rho_{b}(s)$ and $\rho_{b}(r_{x},r_{y})$ can also be acquired by removing the corresponding pre-integrals in~(\ref{Eq_densmatrix_plane}) and normalizing respectively with $\ud \trm{P}_{b}/\ud s$ and $\ud^{2}\trm{P}_{b}/\ud r_{x}\ud r_{y}$. One can thus measure the spin entanglement between the produced pair with different energy in the specified direction by calculating the concurrence $\mC(s)$ and $\mC(r_{x},r_{y})$.

\subsection{Local approximation}
For the NBW process to proceed we of course need to reach the kinematic threshold to produce the (massive) electron-positron pair from the (massless) initial photon.
For laser-particle collisions, where this threshold is reached through the absorption of many photons, we quantify strength of the interaction using the strong-field parameter $\chi$ which must reach $\chi\gtrsim O(1)$.
Sometimes referred to as the `quantum nonlinearity parameter' (because $\chi \propto \hbar$), in a plane-wave electromagnetic background, it takes the simple form $\chi = \xi \eta$. Laser-particle experiments can typically reach $\eta \sim O(0.1)$ and therefore we require $\xi \gg 1$ for NBW to be observable.
In this nonlinear regime, direct calculation using Volkov states is cumbersome and approximate `local' rates are instead derived and employed in numerical simulation of experimental set-ups.

These local rates are inferred by rewriting the double phase integral in \eqnref{Eq_prob_plane} as an integral over the average phase $\vphi=(\phi_{1}+\phi_{2})/2$ and the interference phase $\vtheta=\phi_{2}-\phi_{1}$.
Formally, the limits on the integration in the interference phase are infinite but in many cases a useful approximation can be made by replacing the integration limits with $\pm \Delta\vtheta/2$.
The (phase) region $\Delta\vtheta$ that must be included for an accurate approximation is often referred to as the \emph{formation length} of the process (for more details, see e.g. \cite{Fedotov:2022ely}).
The exponent and pre-exponent in the double phase integral in \eqnref{Eq_prob_plane} are then expanded independently in $\vtheta$ and, which is then integrated over the original limits.
In this, two main expansion approaches are used.
The locally constant field approximation (LCFA) is used when $\xi \gg 1$ since in this limit, the formation length scales as $\Delta\vtheta \sim 1/ \xi$ \cite{ritus85} and hence includes interference effects only on the sub-wavelength scale.
The LCFA corresponds to integrating the `probability rate' for the process occurring in a constant and crossed background, over the local value of the intensity parameter of a non-constant field~\cite{2011PRL035001,2011PRSTAB054401,TangPRA2014,2015PRE023305,TangPRA2019,WAN2020135120,Seipt_2021}.
The range of application of the LCFA has been broadly studied~\cite{Piazza2018PRA012134,blackburn2020radiation}, and efforts have been made to improve its precision~\cite{BenPRA2013,king19a,PiazzaPRA2019,BenPRA042508}.
A second approach, the locally monochromatic approximation (LMA) results in a probability equal to that in a monochromatic background, but with the (constant) amplitude of the potential replaced by the local value of the pulse envelope and the (infinite) phase length factor again replaced by an integral over phase~\cite{LMA063110,Blackburn_2021,blackburn2021higher,TangPRD096019}.
The LMA is less versatile than the LCFA, requiring a laser pulse with a well-defined frequency i.e., $f'(\phi)\ll1$,  but it is valid to much lower values of the intensity and captures harmonic structures in outgoing particle spectra because it includes interference on formation length sales of the order of the wavelength $\Delta\vtheta \sim O(\lambda)$.
(The LMA and LCFA both ignore effects that arise due to intensity gradients in the envelope, but these are generally only important for very short pulses.)
The LMA was used to model the production of NBW pairs in the E144 experiment \cite{bamber99} and is central to modelling the NBW process in the proposed LUXE laser-particle experiment ~\cite{Abramowicz:2021zja,LUXE:2023crk} using the simulation code Ptarmigan \cite{ptarmigan,PoP093903}.
Because of the importance of particle polarisation to entanglement and the fact that the LCFA is sometimes deficient, in describing polarisation effects (e.g. the circular polarisation of a laser~\cite{Tang:2022a,BenPRA2020}), and also because there will be a difference in the entanglement of pairs produced in the low-intensity and high-intensity regime, we will present both these approximations in this work.

The detailed derivation of these approximations have been well presented in~\cite{LMA063110} for LMA and~\cite{ritus85} for LCFA,
Here, the final expression of the density matrices are given below.

\subsubsection{LMA in circularly polarised backgrounds}
After doing the LMA in a circularly polarised laser background and integrating over the positron's transverse momentum, the total probability can be written as~\cite{Tang:2022a,TangPRD096019}
\begin{align}
&\trm{P}_{mc}=\frac{\alpha}{\eta} \int \ud s \int \ud\vphi \sum_{n=\lceil n_{\ast} \rceil}\nonumber\\
&~~~\left[h_{s}\xi^{2}(\vphi)\left(J^{2}_{n+1}+J^{2}_{n-1} - 2J^{2}_{n}\right)+ J^{2}_{n}\right. \nonumber\\
&~~~~\left.  - 4\mathfrak{c}\Gamma_{3} \xi(\vphi) h_{s} \left(\frac{n\eta t s}{r_{n}} - r_{n}\right)J_{n} J'_{n} \right]\,,
\end{align}
where $\xi(\vphi) = \xi_{0}f(\vphi)$, $\lceil n_{\ast} \rceil$ denotes the lowest integer greater than or equal to $n_{\ast}=[1+\xi^{2}(\vphi)]/(2\eta t s)$.
The argument of the Bessel function $J_{n}(\zeta)$ and its derivative $J'_{n}$ is \mbox{$\zeta=\xi(\vphi) r_{n}/(\eta t s)$} and \mbox{$r_{n} =[2n \eta t s -1 - \xi^{2}(\vphi)]^{1/2}$}. The correlated density matrix can be approximated as
\begin{align}
\rho_{b,mc}=&\frac{1}{\trm{P}_{mc}}\frac{\alpha}{\eta} \int \ud s \int \ud\vphi \sum_{n=\lceil n_{\ast} \rceil}\nonumber\\
&~\left[\rho^{0}_{mc} + \Gamma_{1}\rho^{1}_{mc} + \Gamma_{2}\rho^{2}_{mc} + \Gamma_{3}\rho^{3}_{mc} \right]\,,
\end{align}
where
\begin{subequations}
\begin{align}
\rho_{mc}^{0}=&\begin{bmatrix}
\frac{J^{2}_{n}}{4st} & 0 & 0 & 0 \\
0 & h_{s} \mathcal{A}_{mc} + g_{s}\mathcal{R}_{m} &-\mathcal{A}_{mc}/2 & 0\\
0 &-\mathcal{A}_{mc}/2 & h_{s} \mathcal{A}_{mc} - g_{s}\mathcal{R}_{m} & 0\\
0 & 0 & 0 & \frac{J^{2}_{n}}{4st}
\end{bmatrix},\\
\rho^{1}_{mc}=&\begin{bmatrix}
 0 & 0 & 0 &- J^{2}_{n}/(4st) \\
 0 & 0 & 0 & 0 \\
 0 & 0 & 0 & 0 \\
- J^{2}_{n}/(4st) & 0 & 0 & 0
\end{bmatrix},\\
\rho^{2}_{mc}=&\begin{bmatrix}
 0 & 0 & 0 &  iJ^{2}_{n}/(4st)\\
 0 & 0 & 0 & 0 \\
 0 & 0 & 0 & 0 \\
-i J^{2}_{n}/(4st) & 0  & 0  & 0
\end{bmatrix},\\
\rho_{mc}^{3}=&\begin{bmatrix}
\frac{ J^{2}_{n}}{4st} & 0 & 0 & 0 \\
0 &-g_{s}\mathcal{A}_{mc} - h_{s}\mathcal{R}_{m} & \mathcal{R}_{m}/2 & 0\\
0 & \mathcal{R}_{m}/2 & g_{s}\mathcal{A}_{mc} - h_{s}\mathcal{R}_{m} & 0\\
0 & 0 & 0 & \frac{-J^{2}_{n}}{4st}
\end{bmatrix},
\end{align}
\label{Eq_polar_matrix_LMA_C}
\end{subequations}
$\!\!$where
\begin{align}
&\mathcal{R}_{mc}=\mathfrak{c} \frac{\xi^{2}(\vphi)}{2} \left[\frac{1 + \xi^{2}(\vphi)}{n\eta t s} - 1  \right]\left( J^{2}_{n-1} - J^{2}_{n+1}\right)\,,\nonumber\\
&\mathcal{A}_{mc} = \frac{\xi^{2}(\vphi)}{2}\left(J^{2}_{n+1}+J^{2}_{n-1} - 2J^{2}_{n}\right) - J^{2}_{n} \nonumber\,.
\end{align}
\subsubsection{LMA in linearly polarised backgrounds}
In linearly polarised laser fields ($\mathfrak{c}=0$ in Eq.~(\ref{Eq_Com__plane_cir1_slow})), the fast variations in the phase integrals over $\phi_{1,2}$ can be decomposed as
\[\Lambda_{j,n}(\zeta,\beta)=\int_{-\pi}^{\pi}\frac{\ud\phi}{2\pi}\cos^{j}(\phi)e^{i\left[n\phi-\zeta\sin(\phi)+\beta\sin(2\phi)\right]}\,,\]
where $j=0,1,2$, $\Lambda_{j,n}(\zeta,\beta)$ is the generalized Bessel functions~\cite{Reiss1962,landau4} and can be written as the sum of products of the first kind of Bessel functions~\cite{Korsch14947,PRE026707}.

The total probability can be written as
\begin{align}
&\trm{P}_{ml}=\frac{\alpha}{\eta}\int \ud s \int \ud\vphi \int_{0}^{\pi}\frac{\ud\psi}{\pi}\sum_{n=\lceil n_{\ast} \rceil}  \nonumber\\
       &~~\left\{2h_{s}\xi^{2}(\vphi)\left(\Lambda^{2}_{1,n} - \Lambda_{0,n}\Lambda_{2,n}\right) + \Lambda^{2}_{0,n}\right.\nonumber\\
       &~~~\left. -\Gamma_{1}\left[\left(r_{n,x}\Lambda_{0,n} - \xi \Lambda_{1,n}\right)^{2} - r^{2}_{n,y}\Lambda^{2}_{0,n}\right]\right\}\,,
\end{align}
where $n_{\ast}=[1+ \xi^{2}(\vphi)/2]/(2\eta t s)$, $r_{n,x} = r_{n}\cos\psi$, $r_{n,y} = r_{n}\sin\psi$, $r_{n}  =[2n\eta_{\ell} t s - 1 - \xi^{2}(\vphi)/2]^{1/2}$, and the arguments of $\Lambda_{j,n}(\zeta,\beta)$ is given as $\zeta= \xi(\vphi) r_{n}\cos\psi/(\eta st)$ and $\beta=\xi^{2}(\vphi)/(8\eta st)$. The spin correlation density matrix becomes
\begin{align}
\rho_{f,ml}=&\frac{1}{\trm{P}_{ml}}\frac{\alpha}{\eta}\int \ud s \int \ud\vphi \int_{-\pi}^{\pi}\frac{\ud\psi}{2\pi}\sum_{n=\lceil n_{\ast} \rceil}  \nonumber\\
&~\left[\rho^{0}_{ml} + \Gamma_{1}\rho^{1}_{ml} + \Gamma_{2}\rho^{2}_{ml} + \Gamma_{3}\rho^{3}_{ml} \right]\,,
\end{align}
where
\begin{subequations}
\begin{align}
\rho^{0}_{ml}=&\begin{bmatrix}
\frac{\Lambda^{2}_{0,n}}{4st} & 0 & 0 & 0 \\
 0 & h_{s}\mathcal{A}_{ml} &-\frac{1}{2}\mathcal{A}_{ml} & 0 \\
 0 &-\frac{1}{2}\mathcal{A}_{ml} & h_{s}\mathcal{A}_{ml} & 0 \\
 0 & 0 & 0 & \frac{\Lambda^{2}_{0,n}}{4st}
\end{bmatrix}\,,\\
\rho^{1}_{ml}=&\begin{bmatrix}
 0 & 0 & 0 &-\frac{\Lambda^{2}_{0,n}}{4st} \\
 0 & n_{1,m}/2 &-h_{s} n_{1,m} & 0 \\
 0 &-h_{s} n_{1,m} & n_{1,m}/2 & 0 \\
-\frac{\Lambda^{2}_{0,n}}{4st} & 0 & 0 & 0
\end{bmatrix}\,,\\
\rho^{2}_{ml}=& \begin{bmatrix}
 0 & 0 & 0 & i\frac{\Lambda^{2}_{0,n}}{4st}\\
 0 & 0 & -ig_{s} n_{1,m}  & 0\\
 0  & ig_{s} n_{1,m} &   & 0  \\
-i\frac{\Lambda^{2}_{0,n}}{4st} & 0 & 0  & 0
\end{bmatrix},\\
\rho^{3}_{ml}=&\begin{bmatrix}
 \frac{\Lambda^{2}_{0,n}}{4st}  & 0 & 0 & 0 \\
 0 &-g_{s} \mathcal{A}_{ml}  & 0 & 0 \\
 0 & 0 & g_{s} \mathcal{A}_{ml}  & 0 \\
 0 & 0 & 0 & -\frac{\Lambda^{2}_{0,n}}{4st}
\end{bmatrix}\,,
\end{align}
\label{Eq_polar_matrix_LMA_L}
\end{subequations}
$\!\!$where
\begin{align}
&\mathcal{A}_{ml} = \xi^{2}(\vphi)\left(\Lambda^{2}_{1,n} - \Lambda_{0,n}\Lambda_{2,n}\right) - \Lambda^{2}_{0,n}\,,\nonumber\\
&n_{1,m} =r^{2}_{n,y}\Lambda^{2}_{0,n} - \left[r_{n,x}\Lambda_{0,n} - \xi(\vphi) \Lambda_{1,n}\right]^{2}\nonumber
\end{align}
\subsubsection{Locally constant field approximation}
The LCFA is performed by expanding the pre-exponent in the interference phase $\vtheta=\phi_{1}-\phi_{2} $ to second order and the exponent to third order. This means the following replacements are made for the potential in the pre-exponent:
\begin{align}
a^{\mu}(\phi_{1})&=a^{\mu}(\varphi)+a'^{\mu}(\varphi)\vtheta/2\,,\nonumber\\
a^{\mu}(\phi_{2})&=a^{\mu}(\varphi)-a'^{\mu}(\varphi)\vtheta/2\,,\nonumber
\end{align}
where $\vphi=(\phi_{1}+\phi_{2})/2$ is the averaged phase, the derivative of vector potential $a'^{\mu}(\varphi)=-m[0,\xiE_{x}(\vphi),\xiE_{y}(\vphi),0]$ gives the laser's electric field $\bm{\xiE}=(\xiE_{x},\xiE_{y})$.

After the integral over the positron's transverse momentum, the total probability can be written as~\cite{Tang:2022a}
\begin{align}
\trm{P}_{b,c}=&\frac{\alpha}{\eta } \int\ud s \int \ud \vphi \nonumber\\
&~\left[ \textrm{Ai}_{1} - \textrm{Ai}'/z\left(4h_{s}-\Gamma_{1} \vsigma_{1} -\Gamma_{2} \vsigma_{2} \right) \right]\,,
\label{Eq_LCFA_Spectrum}
\end{align}
where $\vsigma_{1}(\vphi) = (\xiE^2_{x} -\xiE^2_{y})/|\bm{\xiE}|^2$ and $\vsigma_{2}(\vphi) = 2\xiE_{x}\xiE_{y}/|\bm{\xiE}|^2$ denote the linear polarisation of the local field and $\textrm{Ai}'\equiv \trm{Ai}'(z)$ is the derivative of the Airy function $\trm{Ai}(z)$ with the argument $z=(s t \eta |\pmb{\xiE}|)^{-2/3}$ and $\trm{Ai}_{1}\equiv \trm{Ai}_{1}(z) = \int^{\infty}_{z}\ud x \trm{Ai}(x)$.
The final density matrix can then be given as
\begin{align}
&\rho_{b,c}
=\frac{1}{\trm{P}_{b,c}}\frac{\alpha}{\eta} \int\ud s \int \ud\vphi \nonumber\\
&~~~~~~~~~~~~~~~\left(\rho^{0}_{c} + \Gamma_{1}\rho^{1}_{c} + \Gamma_{2}\rho^{2}_{c} + \Gamma_{3}\rho^{3}_{c} \right)\,,
\end{align}
where
\begin{subequations}
\begin{align}
\rho^{0}_{c}=&
\begin{bmatrix}
 \frac{\textrm{Ai}_{1}}{4 st} & \frac{i\tilde{\xiE}^{*}\textrm{Ai}}{4s \sqrt{z}} &-\frac{i\tilde{\xiE}^{*} \textrm{Ai}}{4t \sqrt{z}} & 0 \\
 \frac{-i\tilde{\xiE}\textrm{Ai}}{4s \sqrt{z}} & h_{s}\mathcal{A}_{c} &-\mathcal{A}_{c}/2 &\frac{-i\tilde{\xiE}^{*}\textrm{Ai}}{4t \sqrt{z}} \\
 \frac{ i\tilde{\xiE}\textrm{Ai}}{4t \sqrt{z}} &-\mathcal{A}_{c}/2 & h_{s}\mathcal{A}_{c} &\frac{ i\tilde{\xiE}^{*}\textrm{Ai}}{4s \sqrt{z}} \\
 0 &\frac{i\tilde{\xiE}\textrm{Ai}}{4t \sqrt{z}} &\frac{-i\tilde{\xiE}\textrm{Ai}}{4s \sqrt{z}} & \frac{\textrm{Ai}_{1}}{4 st}
\end{bmatrix}\,,\\
\rho^{1}_{c}=&\begin{bmatrix}
 0 & \frac{-i\tilde{\xiE}\textrm{Ai}}{4t\sqrt{z}} & \frac{i\tilde{\xiE}\textrm{Ai}}{4s\sqrt{z}} &-\frac{\textrm{Ai}_{1}}{4st} \\
 \frac{ i\tilde{\xiE}^{*}\textrm{Ai}}{4t\sqrt{z}} & \frac{1}{2}\vsigma_{1}\frac{\trm{Ai}'}{z} & n_{12} \frac{\trm{Ai}'}{z} & \frac{ i\tilde{\xiE}\textrm{Ai}}{4s\sqrt{z}}\\
 \frac{-i\tilde{\xiE}^{*}\textrm{Ai}}{4s\sqrt{z}} & n^{*}_{12} \frac{\trm{Ai}'}{z} & \frac{1}{2}\vsigma_{1}\frac{\trm{Ai}'}{z}  & \frac{-i\tilde{\xiE}\textrm{Ai}}{4t\sqrt{z}}\\
-\frac{\textrm{Ai}_{1}}{4st} & \frac{-i\tilde{\xiE}^{*}\textrm{Ai}}{4s\sqrt{z}} &\frac{i\tilde{\xiE}^{*}\textrm{Ai}}{4t\sqrt{z}} & 0
\end{bmatrix},\\
\rho^{2}_{c}=&\begin{bmatrix}
 0 &\frac{-\tilde{\xiE}\textrm{Ai}}{4t\sqrt{z}}  & \frac{\tilde{\xiE}\textrm{Ai}}{4s\sqrt{z}} & \frac{i\textrm{Ai}_{1}}{4st}\\
\frac{-\tilde{\xiE}^{*}\textrm{Ai}}{4t\sqrt{z}}  & \frac{1}{2}\vsigma_{2}\frac{\trm{Ai}'}{z}  & n_{21}\frac{\trm{Ai}'}{z} & \frac{\tilde{\xiE}^{*}\textrm{Ai}}{4s\sqrt{z}}\\
 \frac{\tilde{\xiE}^{*}\textrm{Ai}}{4s\sqrt{z}}  & n^{*}_{21} \frac{\trm{Ai}'}{z} & \frac{1}{2}\vsigma_{2}\frac{\trm{Ai}'}{z}  & \frac{-\tilde{\xiE}^{*}\textrm{Ai}}{4t\sqrt{z}}  \\
 \frac{-i\textrm{Ai}_{1}}{4st} & \frac{\tilde{\xiE}^{*}\textrm{Ai}}{4s\sqrt{z}}  &\frac{-\tilde{\xiE}^{*}\textrm{Ai}}{4t\sqrt{z}}  & 0
\end{bmatrix},\\
\rho^{3}_{c}=&\begin{bmatrix}
 \frac{\textrm{Ai}_{1}}{4st}  & \frac{i\tilde{\xiE}^{*}\textrm{Ai}}{4s\sqrt{z}} & \frac{-i\tilde{\xiE}^{*}\textrm{Ai}}{4t\sqrt{z}} & 0 \\
 \frac{-i\tilde{\xiE}\textrm{Ai}}{4s\sqrt{z}} &-g_{s}\mathcal{A}_{c} & 0 & \frac{ i\tilde{\xiE}^{*}\textrm{Ai}}{4t\sqrt{z}} \\
 \frac{ i\tilde{\xiE}\textrm{Ai}}{4t\sqrt{z}} & 0 & g_{s}\mathcal{A}_{c} & \frac{-i\tilde{\xiE}^{*}\textrm{Ai}}{4s\sqrt{z}} \\
 0 & \frac{-i\tilde{\xiE}\textrm{Ai}}{4t\sqrt{z}} & \frac{i\tilde{\xiE}\textrm{Ai}}{4s\sqrt{z}} &-\frac{\textrm{Ai}_{1}}{4st}
\end{bmatrix}\,,
\end{align}
\label{Eq_polar_matrix_LCFA}
\end{subequations}
$\!\!$with $\tilde{\xiE} = (\xiE_{x} + i\xiE_{y})/|\bm{\xiE}|$, $\mathcal{A}_{c} = -2\trm{Ai}'(z)/z - \trm{Ai}_{1}(z)$, $n_{12} = ig_{s}\vsigma_{2} - h_{s}\vsigma_{1}$ and $n_{21} =  -(ig_{s} \vsigma_{1} + h_{s} \vsigma_{2})$.
Note that the general expression in (\ref{Eq_polar_matrix}c) has the $\rho^3_{b}$ trace $\sim\mathcal{R}$, and
the tracelessness of $\rho^3_{c}$ is an early indication that the LCFA will not be able to resolve the effect of photon helicity on the concurrence.

\section{Numerical results}~\label{Sec_4}
In this section, we present results of our numerical investigation into how the spin entanglement of the electron and positron produced by the nonlinear Breit-Wheeler process varies when the electromagnetic background and incident photon properties are varied.
For the background, we vary the intensity of the plane wave pulse and consider circular and linear polarisation; for the incident photon, we vary the energy parameter and polarisation degree. The degree of spin entanglement is assessed by calculating the concurrence, $0\leq \mathcal{C} \leq 1$. We look at two cases: in \sxnref{sec:3.1a} the intensity and energy dependence is investigated for monoenergetic photons with the fixed polarisation; in \sxnref{NLC_photon} the photon is replaced by a distribution in energy and polarisation generated by a quasi-linear Compton source. In both cases we assume an $N$-cycle plane wave pulse envelope of the form: $f(\phi)=\cos^{2}(\phi/2N)$ for $|\phi|<N\pi$ with $f(\phi)=0$ otherwise, and choose $N=4$, corresponding to a full-width-at-half-maximum pulse duration of  $5.33~\trm{fs}$. (The short duration is chosen for ease of calculation; the results will not be significantly affected for a longer pulse.) We assume a laser carrier frequency of $1.55\,\trm{eV}$. Because the intensities we consider span the full range from perturbative $\xi \ll 1$ to the non-perturbative $\xi\gg1$ in the charge-field coupling, we calculate using the LCFA, the LMA, and the direct evaluation of the full QED expressions without local approximation.

\subsection{Monoenergetic photons}\label{sec:3.1a}
We consider two set-ups: i) the laser is circularly polarised and the photon polarisation is described using helicity eigenstates of the polarisation operator quantified by the $\Gamma_{3}$ Stokes parameter; ii) the laser is linearly polarised and the photon polarisation is described using a linear polarisation basis and the $\Gamma_{1}$ Stokes parameter.

First, the variation of concurrence with the created electron and positron lightfront momentum (an approximate measure of their energy) is presented for an incident photon with energy parameter $\eta=0.2$ and laser intensity  $\xi=2$.
We see, the concurrence spectrum $\mC(s)$ is symmetric at $s=0.5$ just like the probability spectrum $\ud\trm{P}_{b}/\ud s$, which follows due to the underlying charge-parity conjugation (CP) symmetry of QED.
In the central region around $s=0.5$, where the energy difference between the pair particles is small (i.e. $|s-t|\to 0$), the electron and positron are created with maximum value of concurrence and probability. In contrast, as the difference in energy parameter increases
 (i.e. $|s-t|\to 1$), the concurrence decreases, as does the creation probabity as well.

Second, the energy parameter of the photon fixed at $\eta=0.2$ and the effect on the concurrence of varying the laser intensity is calculated.
Three regions of behaviour can be identified.
In the perturbative regime $\xi\ll1$, the concurrence is independent of $\xi$. This is because in this limit all non-zero elements of the unnormalised density matrix are proportional to $\xi^{2}$ and so too is the total probability that the density matrix is normalised with, hence cancelling the dependence on $\xi$.
Because pair-creation is only energetically accessible at low $\xi$ by absorbing a large momentum from the variation of the pulse envelope, the perturbative limit is not reproduced by the local approximations.
(In experiments, the perturbative limit can only realistically be reached when the energy parameter $\eta$, which depends on the carrier frequency of the EM field, satisfies $\eta \gtrsim 2$. In our direct QED calculations, the perturbative limit can be reached in another way, due to very high frequency components being included from the pulse envelope. However this is unrealistic since those components would not be transmitted by optical elements in a real experiment.) In the intermediate intensity regime $\xi \sim O(1)$ the formation length is of the order of a wavelength, and so the LMA agrees very well with the full QED calculation. In the high intensity regime $\xi \gg 1$, the formation length is sub-wavelength and the LCFA becomes accurate.

Third, the intensity is fixed at $\xi=2$ and the energy parameter of the photon is varied, which changes the linearity of the process. In the low-energy regime, which for this value of $\xi$ is identified from the results as $\eta < 0.04$, (for a laser with frequency $1.55\,\trm{eV}$, this corresponds to photon energies lower than $3.3\,\trm{GeV}$),  pair-creation is linear.
This is because the centre-of-mass energy is so low that the non-perturbative contribution from absorbing many laser photons required to reach the pair-creation threshold is suppressed more strongly than the Linear Breit-Wheeler (LBW) process that takes high frequencies from the pulse envelope (see e.g.~\cite{TangPRD096019} for more detail).
As the photon energy is increased to an intermediate regime, the non-perturbative contribution becomes dominant and the process is highly nonlinear. As the energy is further increased, eventually the high energy regime is reached, in which only a few harmonics, i.e. \emph{net} number of laser photons, are required for pair creation to proceed.
The threshold for pair-creation from the $n$th harmonic to proceed is:
\begin{align}
\eta>\frac{2(1+\upsilon\xi^{2})}{n}
\label{Eq_harmonic_edges}
\end{align}
where $\upsilon=1/2$ for a linearly-polarised background and $\upsilon=1$ for a circularly polarised background). The opening of the $n=2$ and $n=1$ harmonics are normally particularly clear in spectra. Even though only one or two laser photons are required for pair-creation to proceed at these energies, since $\xi \not \ll 1$ here, many laser photons contribute and the harmonic is the \emph{net} number of laser photons. i.e. the process is still highly nonlinear.

\begin{figure}[t!!!]
\center{\includegraphics[width=0.48\textwidth]{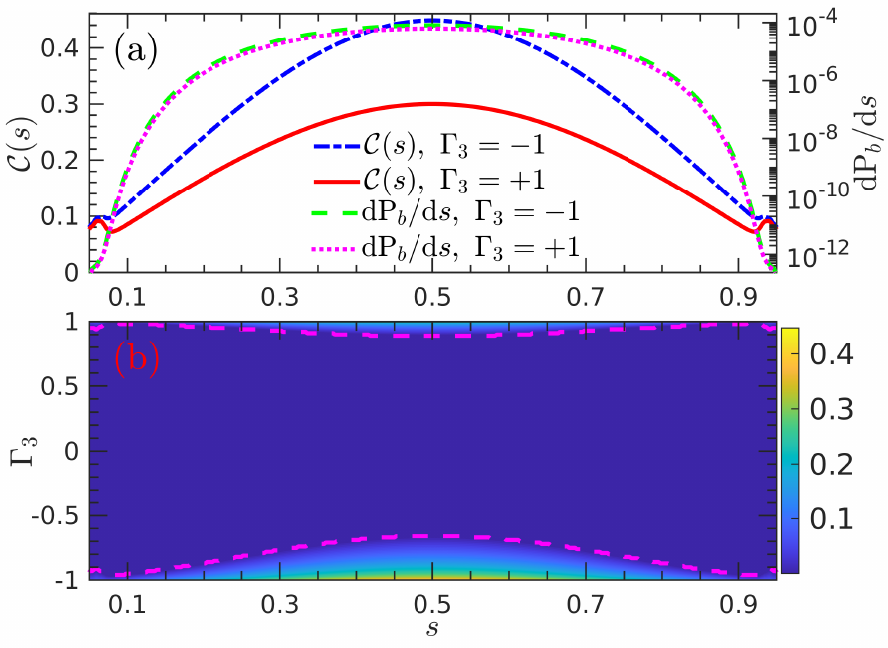}}
\caption{The electron-positron spin entanglement's dependence on the lightfront momentum of positron in a $4$-cycle, circularly polarised ($\maf{c}=+1$) laser background with intensity $\xi=2$ that collides head-on with photons of energy parameter $\eta=0.2$.  Fig. (a) contains the dependence of fully polarised photons with $\Gamma_{3}=\pm1$ and Fig. (b) for partially polarised photons with $-1<\Gamma_{3}<1$.
The probability spectra from fully-polarised photons with $\Gamma_{3}=\pm1$ are also given in (a) using the right-hand vertical axis to indicate the dominant region for pair-creation.
The magenta dashed line in (b) indicates a region in which $\mC(s)=0$.}
\label{Fig_spec_cir}
\end{figure}

\emph{Circularly polarised laser background -}~
The concurrence spectrum $\mC(s)$ is plotted in Fig.~\ref{Fig_spec_cir} for pair-creation in a circularly-polarised background with $\mathfrak{c}=+1$ (see \eqnref{Eq_Com__plane_cir1_slow}).
In Fig.~\ref{Fig_spec_cir}~(a) we see that for both helicity eigenstates $\Gamma_{3}=\pm1$ the concurrence is at its largest when the created electron and positron have the same energy ($s=0.5$), and decreases as $s$ goes to $0$ or $1$.
In these limits $s\rightarrow0,1$ we see a modulation in the concurrence due to pulse-envelope effects. This effect can also be seen in the probability spectra (right-hand axis) as the decrease in probability away from $s=0.5$ slows down. We can also see that photons in a helicity state with $\Gamma_{3}=-\mathfrak{c}$, i.e. opposite to the laser rotation, create electron-positron pairs with a stronger spin entanglement than those created by photons with $\Gamma_{3}=+\mathfrak{c}$, i.e. parallel to the laser polarisation.
Although difficult to see due to the log scale on the plot, photons with $\Gamma_{3}=-\mathfrak{c}$ also have a higher probability of creating pairs than those with $\Gamma_{3}=+\mathfrak{c}$~\cite{TangPRD056003}. The variation of the concurrence spectrum $\mC(s)$ with the photon polarisation is plotted in Fig.~\ref{Fig_spec_cir} (b).
We see that to have strong spin entanglement we require the incident photon to be highly polarised with $|\Gamma_{3}|\approx 1$.
Away from this limit the entanglement quickly decreases to zero, with the magenta dashed lines in Fig.~\ref{Fig_spec_cir} (b) outlining regions where the concurrence is exactly zero.

\begin{figure}[t!!!]
\center{\includegraphics[width=0.48\textwidth]{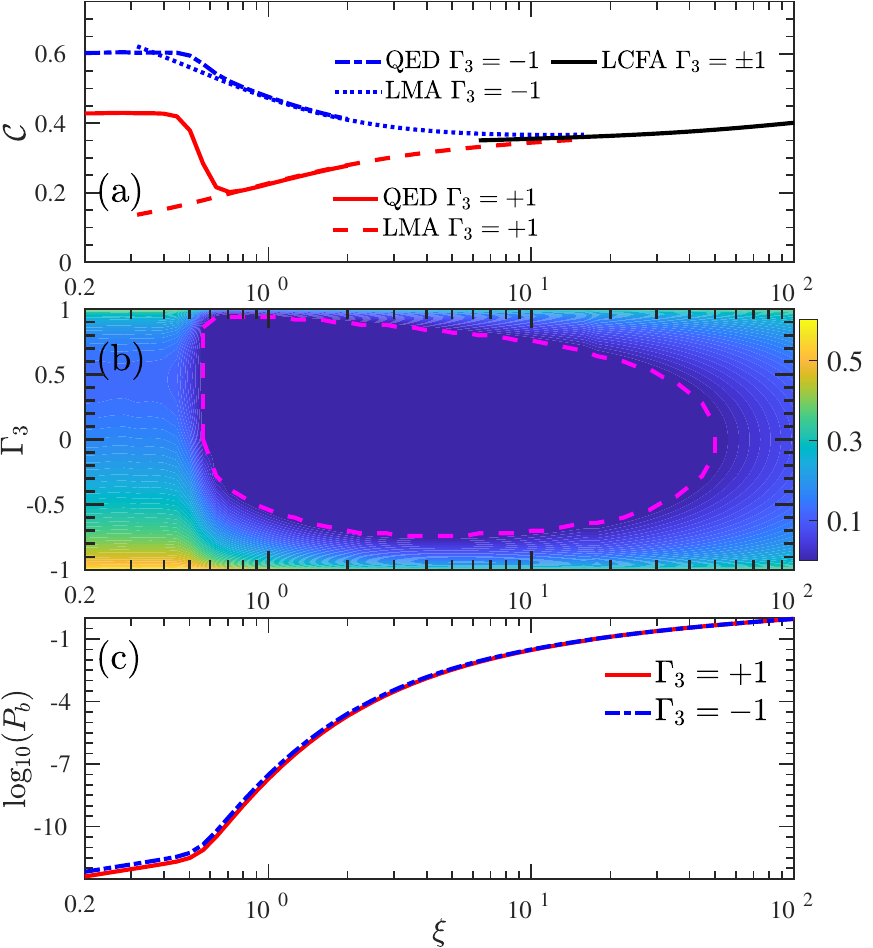}}
\caption{The electron-positron spin entanglement's dependence on the intensity $\xi$ of a $4$-cycle, circularly polarised ($\maf{c}=+1$) laser background that collides head-on with photons of energy parameter $\eta=0.2$.  Fig. (a) contains the dependence of fully polarised photons with $\Gamma_{3}=\pm1$ and fig. (b) for partially polarised photons with $-1<\Gamma_{3}<1$. The numerical data for $0.2<\xi<2$ is from evaluating the full QED expression; for $2.0<\xi<15.8$ from the LMA and $\xi>15.8$ from the LCFA.
Fig. (c) is a plot of the total probability where the perturbative contribution from the pulse envelope is clearly visible for $\xi\lesssim 0.5$.
The magenta dashed line in Fig.~(b) indicates a region in which $\mC=0$.}
\label{Fig1_xi_cir}
\end{figure}

\begin{figure}[t!!!]
\center{\includegraphics[width=0.48\textwidth]{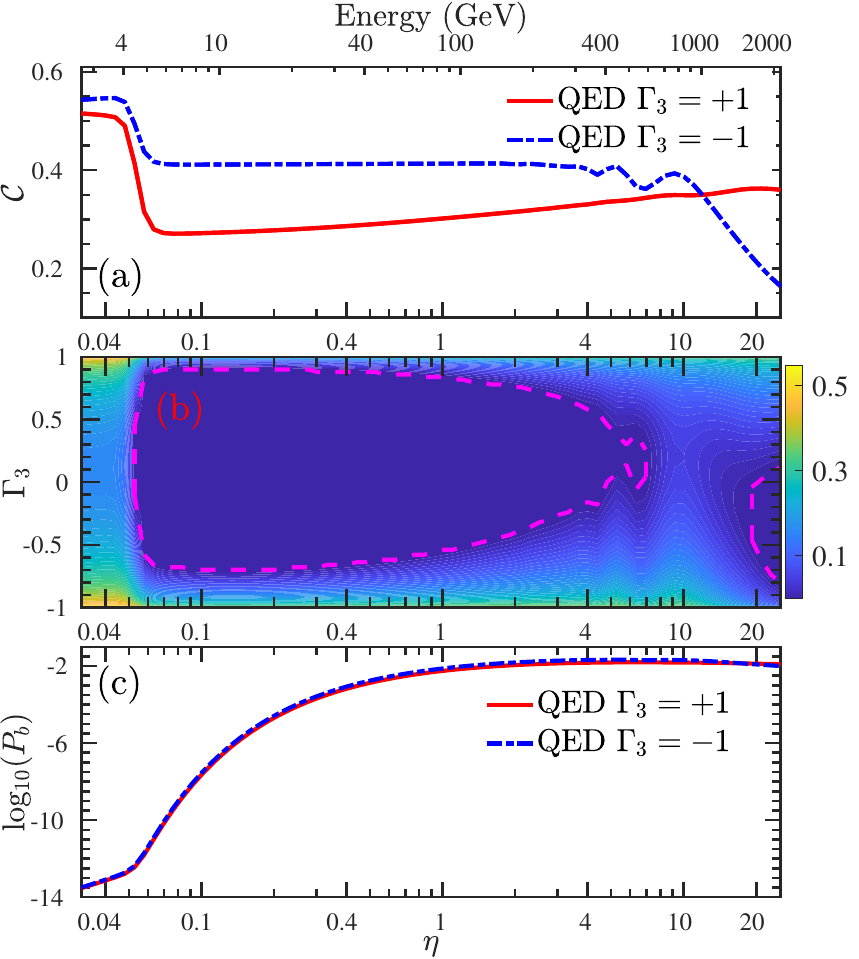}}
\caption{The electron-positron spin entanglement's dependence on the photon energy parameter $\eta$ in a $4$-cycle, circularly polarised ($\maf{c}=+1$) laser background with $\xi=2$.  Fig. (a) contains the dependence of fully polarised photons with $\Gamma_{3}=\pm1$ and fig. (b) for partially polarised photons with $-1<\Gamma_{3}<1$.
Fig. (c) is a plot of the total probability where the perturbative contribution from the pulse envelope is clearly visible, here for $\eta\lesssim 0.05$. The collision energy parameter $\eta$ changes from $0.03$ to $25$, with the corresponding change in photon energy (for a laser frequency of $1.55\,\trm{eV}$), from $2.7~\trm{GeV}$ to $100~\trm{GeV}$, shown on the top axis in (a). The magenta dashed line in Fig.~(b) indicates a region in which $\mC=0$.}
\label{Fig3_eta_cir}
\end{figure}

Fig.~\ref{Fig1_xi_cir} shows the variation of the concurrence with the intensity in a circularly-polarised background with $\mathfrak{c}=+1$.
Fig.~\ref{Fig1_xi_cir}(a) shows how the concurrence varies for $\Gamma_{3}=\pm1$, with the curves split into regions where different approximations are used in the calculations. In the perturbative and intermediate intensity regimes the concurrence is clearly plotted as two distinct curves; however, as the intensity is increased, these curves converge. This behaviour can be explained by the formation length becoming so short at high intensities that the pair is formed instantaneously, therefore being insensitive to the direction of rotation of photon polarisation, which varies on a lengthscale of the order of a wavelength.
It is noteworthy that the LCFA does not become accurate until $\xi \gtrsim 10$, i.e. a slightly higher intensity than usual when evaluating the scattering probability~\cite{Tang:2022a}.
Overall, we can see that the photons with $\Gamma_{3}=-\mathfrak{c}$ create electron-positron pairs with a stronger spin entanglement than created by photons with $\Gamma_{3}=+\mathfrak{c}$. In Fig.~\ref{Fig1_xi_cir} (b), the behaviour of the concurrence is plotted for partially-polarised photons.
The strongest entanglement requires the incident photon to be highly polarised with $|\Gamma_{3}|\approx 1$.
With decreasing polarisation $|\Gamma_{3}|$, the entanglement decreases quickly to zero; see the region surrounded by the magenta dashed line in Fig.~\ref{Fig1_xi_cir} (b). For completely unpolarised photon $\Gamma_{3}=0$,  non-zero concurrence can only appear in the perturbative regime, in which $\mC=0.18$, and in the region of $\xi = 10^{2}$ in which $\mC<0.1$.

The variation of the concurrence with the photon energy $\eta$ is presented in Fig.~\ref{Fig3_eta_cir}.
In the low-energy, perturbative regime where $\eta<0.04$, the concurrence becomes independent of the photon energy.
In the plot for fully polarised photons in Fig.~\ref{Fig3_eta_cir} (a), as the photon energy is increased from this low-energy regime, the concurrence quickly falls, and then varies only slowly with $\eta$ in the intermediate energy regime ($\eta \sim O(1)$).
Here, as in the case of varying the intensity, the spin entanglement created by the photon with $\Gamma_{3}=-\mathfrak{c}$ polarised perpendicular to the laser polarisation, is stronger than that created by the photon with the parallel polarisation $\Gamma_{3}=+\mathfrak{c}$. As the energy is further increased to the harmonic regime, the concurrence clearly oscillates around the second ($\eta=5$) and first ($\eta=10$) harmonic {edges} for photons polarised with $\Gamma_{3}=-\mathfrak{c}$. At energies above this, the order of the concurrence switches and the concurrence is highest for pairs created by photons in the $\Gamma_{3}=+\mathfrak{c}$ state.

\begin{figure}[t!!!]
\center{\includegraphics[width=0.48\textwidth]{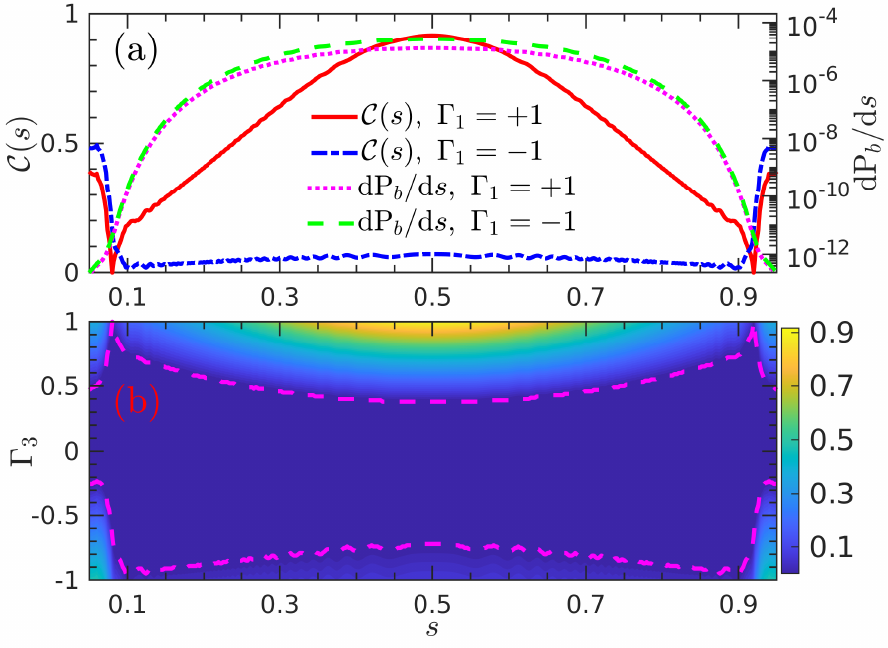}}
\caption{The electron-positron spin entanglement's dependence on the lightfront momentum of positron in a $4$-cycle, linearly polarised laser background with intensity $\xi=2$ that collides head-on with photons of energy parameter $\eta=0.2$.
Fig.~(a) contains the dependence of fully polarised photons with $\Gamma_{1}=\pm1$ and fig. (b) for partially polarised photons with $-1<\Gamma_{1}<1$.
The probability spectra from fully polarised photons with $\Gamma_{1}=\pm1$ are also given in (a) with the right-hand vertical axis to indicate the dominant region for pair-creation.
The magenta dashed line in (b) indicates a region in which $\mC(s)=0$.}
\label{Fig_spec_lin}
\end{figure}

\emph{Linearly polarised laser background} -
The concurrence spectrum $\mC(s)$ is plotted in Fig.~\ref{Fig_spec_lin} for pair-creation in a linearly-polarised laser.
In Fig.~\ref{Fig_spec_lin}~(a) we see that similar to the circularly-polarised case, the degree of entanglement between the particles is largest when they have the same energy and decreases as their energy difference grows.
However in this case we see that at the limits $s\to 0,1$ the concurrence increases sharply.
In contrast to the circularly-polarised case, photons with polarisation parallel to the laser polarisation ($\Gamma_{1}= +1$) create pairs with much larger concurrence than those created by photons polarised perpendicular to the laser polarisation ($\Gamma_{1}= -1$).
For the parallel polarisation case we find $\mC(0.5)=0.92$ while for perpendicular polarised case we find $\mC(0.5)=0.07$. Any significant degree of entanglement requires the incident photon to be highly polarised with $\Gamma_{1}\approx 1$, with the magenta dashed lines in Fig.~\ref{Fig_spec_lin} (b) again outlining regions where the concurrence is exactly zero.

\begin{figure}[h!!]
\center{\includegraphics[width=0.48\textwidth]{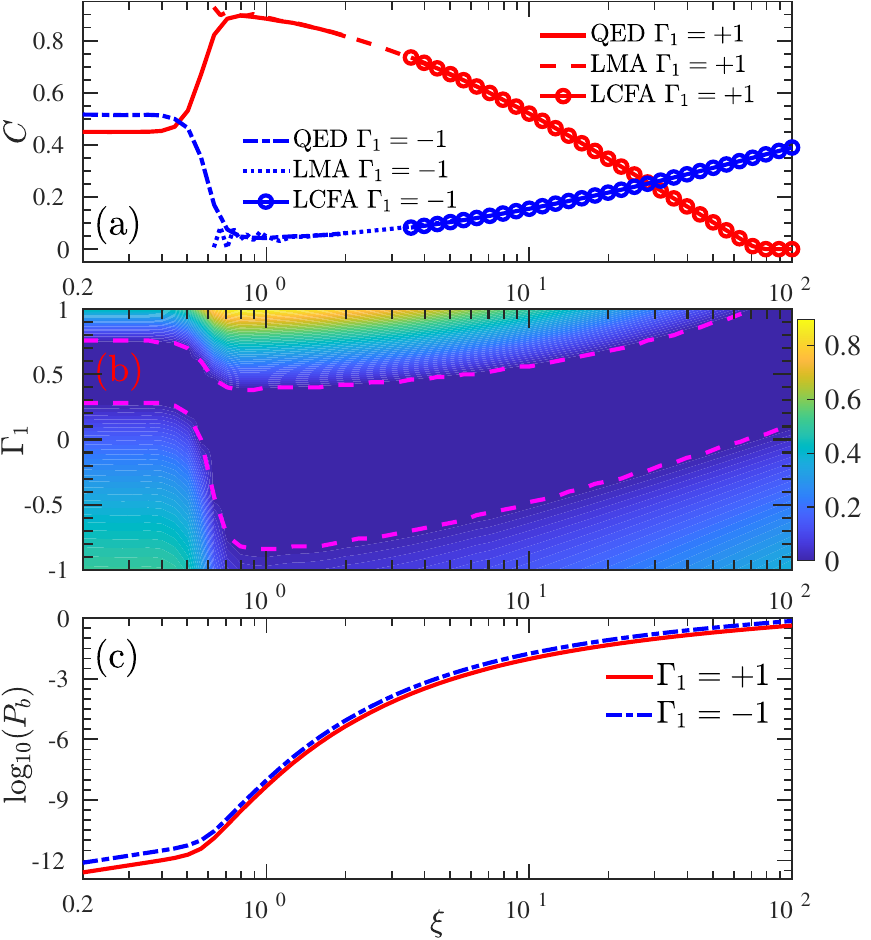}}
\caption{The electron-positron spin entanglement's dependence on the intensity $\xi$ of a $4$-cycle, linearly polarised laser background that collides  with photons of energy parameter $\eta=0.2$.  Fig. (a) contains the dependence of fully polarised photons with $\Gamma_{1}=\pm1$ and fig. (b) for partially polarised photons with $-1<\Gamma_{1}<1$. Fig. (c) is a plot of the total probability where the perturbative contribution from the pulse envelope is clearly visible for $\xi\lesssim 0.6$.
The numerical data for $0.2<\xi<2$ is from evaluating the full QED expression; for $2.0<\xi<15.8$ from the LMA and $\xi>15.8$ from the LCFA.
The magenta dashed line in Fig.~(b) indicates a region in which $\mC=0$.}
\label{Fig2_xi_lin}
\end{figure}

The variation of the spin entanglement with intensity $\xi$ of a linearly-polarised background is shown in Fig.~\ref{Fig2_xi_lin}.
Three different regimes can be identified, just as in the case of a circularly-polarised background, however there are important differences.
In the plot for the two photon polarisation eigenstates, \figrefa{Fig2_xi_lin}, it can be seen that the concurrence in the perturbative regime is similar to the circularly-polarised case (the concurrence $\mC=0.52$ for $\Gamma_{1}=-1$ and $\mC=0.45$ for $\Gamma_{1}=+1$).
However, as the intensity is increased to the intermediate regime, the concurrence varies significantly and actually \emph{increases} in the $\Gamma_{1}=+1$ eigenstate to a maximum of $\mC=0.9$ at $\xi=0.8$ whilst the concurrence falls in the $\Gamma_{1}=-1$ eigenstate.
In the high intensity regime, the relative entanglement of the two polarisation eigenstates switches sign; the concurrence of the $\Gamma_{1}=+1$ decreases monotonically (at $\xi=100$, $\mC=0$) whereas the concurrence of the $\Gamma_{1}=-1$ eigenstate increases monotonically (at $\xi=100$, $\mC\approx 0.39$).
Similar to the circularly polarised background case, the strongest entanglement is created with the highly polarised photons $|\Gamma_{1}|\approx 1$, as shown in Fig.~\ref{Fig2_xi_lin} (b).
However, different to the circular polarisation case, there is an island of high entanglement at $0.5\lesssim\xi\lesssim5$ for photons mostly polarised in the $\Gamma_{1}=+1$ eigenstate, providing an example of how a highly entangled pair can be created in the non-perturbative $\xi \not \ll 1$ regime.

\begin{figure}[t!!!]
\center{\includegraphics[width=0.48\textwidth]{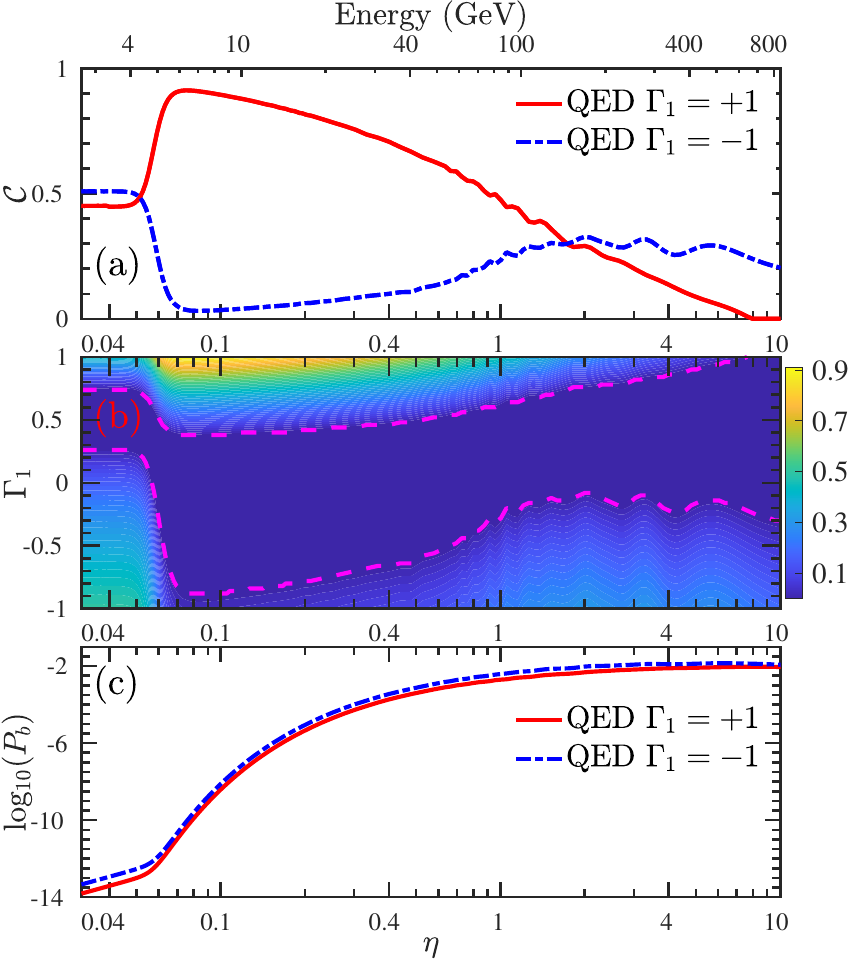}}
\caption{The electron-positron spin entanglement's dependence on the photon energy parameter $\eta$ in a $4$-cycle, linearly-polarised laser background with $\xi=2$.  Fig. (a) contains the dependence of fully polarised photons with $\Gamma_{1}=\pm1$ and fig. (b) for partially polarised photons with $-1<\Gamma_{1}<1$.
Fig. (c) is a plot of the total probability where the perturbative contribution from the pulse envelope is clearly visible for $\eta\lesssim 0.06$. The collision energy parameter $\eta$ changes from $0.03$ to $10$, with the corresponding change in photon energy, from $2.7~\trm{GeV}$ to $100~\trm{GeV}$, shown on the top axis in (a). The magenta dashed line in Fig.~(b) indicates a region in which $\mC=0$.}
\label{Fig4_eta_lin}
\end{figure}

The variation of concurrence with energy for fixed intensity $\xi=2$ is shown in Fig.~\ref{Fig4_eta_lin}, where again a perturbative $\eta \lesssim 0.05$, an intermediate and a high-energy regime can be noted.
In fact, there is a strong similarity with the intensity plots in \figref{Fig2_xi_lin}.
An important difference in the linearly polarised case can be seen in Fig.~\ref{Fig4_eta_lin} (a); in the intermediate energy regime the concurrence changes markedly from the linear regime and \emph{increases} for the photon polarisation $\Gamma_{1}=+1$, parallel to the laser field, while decreasing for the perpendicular polarisation $\Gamma_{1}=-1$. The maximum concurrence of approximately $\mC=0.91$ is hence found at $\eta=0.075$, corresponding to $6.4~\trm{GeV}$ for a laser frequency of $1.55\,\trm{eV}$.
As the energy parameter is increased still further into the high energy regime, oscillation in the concurrence for the second ($\eta=3$) and first ($\eta=6$) harmonic can be clearly noted in $\Gamma_{1}=-1$ polarisation state.
We note, just as in the circularly-polarised case, the order of which photon polarisation state leads to the largest concurrence again switches in the high energy regime, although both components have a low value $\lesssim 0.3$ of the concurrence.

We conclude this section by reiterating that the highest level of spin-entanglement of the electron-positron pair can be found in the intermediate intensity regime for a highly polarised source of photons. An example scheme to generate these photons is given in the next section.

\subsection{Compton photon source}~\label{NLC_photon}
In this section, we consider the scattering of an initially unpolarised electron beam with a laser pulse to produce spin-entangled electron-positron pairs. The laser pulse can split into a double-pulse that act in two stages: I) the first pulse colliding via nonlinear Compton scattering (NLC) with the electron beam to produce a polarised photon source \cite{tang2020highly,NPAllopticalNLC}; II) the second pulse colliding with the photons (after filtering away any charges with magnets) to produce pairs, whose spin is then measured downstream after further separation with magnets. For the electron and laser beam parameters, we take those typical of upcoming experments such as LUXE~\cite{Abramowicz:2021zja, LUXE:2023crk,LUXE:2025wuo} and E320~\cite{chen22} operating with electron energies $\sim O(10)~\trm{GeV}$ ($\eta \sim O(0.1)$) and laser intensities $\xi\sim O(1)$.

In stage I) nonlinear Compton  produces photons most abundantly in the same polarisation state as that of the laser pulse~\cite{serbo04,BenPRA2020}.
For the parameters of interest, it can be seen from Figs.~\ref{Fig1_xi_cir} and~\ref{Fig3_eta_cir} that spin-entanglement is maximised in a circularly polarised background if the photon is \emph{perpendicular} to the laser beam ($\Gamma_{3}=-\mathfrak{c}$) but from Figs.~\ref{Fig2_xi_lin} and~\ref{Fig4_eta_lin} maximised in a linearly polarised background if the photon is polarised \emph{parallel} to the laser beam. This suggests two experimental scenarios: one with two circularly polarised pulses rotating in opposite directions (which we will refer to as the `CP-setup') and one with two pulses linearly polarised in the same direction (the `LP-setup').
The polarisation degree of the Compton photon beam can be finely controlled by applying an angular cut with a finite acceptance $\Delta \theta$ downstream of the electron beam in stage I)~\cite{tang2020highly}. This finite angular acceptance $\Delta\theta \propto w_{l}/d$ can be realized in experiments by simply adjusting the distance $d$ between the double pulses, where $w_{l}$ indicates the transverse waist of the second pulse.
The photon beam is sufficiently narrow that variations in the energy parameter, $\eta\propto 1+\cos(\Delta\theta/2)$, are insignificant for the NBW pair creation process downstream.
(For the parameters we consider, it has been shown \cite{Tang2022_096004} that changing the relative polarisation of the two laser pulses can result in a change to the final pair yield of about $20\%$.)

The process of nonlinear Compton scattering has been broadly discussed in the literature~\cite{harvey09,PRA022101,seipt2017volkov,MRE0196125}.
Here, a similar approach to ~\cite{tang2020highly} is used for calculatig the spectrum and polarisation of the emitted photon beam within a specified angular acceptance, see Appendix.~\ref{App_NLC} for an introduction.
Let the Compton photon beam energy spectrum be $\ud \trm{P}_{c}/\ud s_{c}$ with polarisation $\bm{\Gamma}_{c}(s_{c})$, where $s_{c}=\eta/\eta_{p}$ is fraction of lightfront momentum taken by the emitted photon from the parent electron and $\eta_{p}=k\cdot p/m^{2}$ is the energy parameter of the electron. Then the probability to produce an electron-positron pair can be written as
\begin{align}
\trm{P}_{t} = \int^{1}_{0}\ud s_{c} \frac{\ud \trm{P}_{c}}{\ud s_{c}} \int^{s_{c}}_{0} \frac{\ud s_{b}}{s_{c}} \frac{\ud}{\ud s}\trm{P}_{b}[s;\eta,\bm{\Gamma}_{c}(s_{c})]\,,
\end{align}
where $s_{b} = k\cdot q/k\cdot p$ denotes the fraction of light-front momentum taken by the created positron from the parent electron, and the fraction parameter $s$ can now be given as $s=s_{b}/s_{c}$. The final spin-correlation density matrix can be written as
\begin{align}
\rho_{t} = \frac{1}{\trm{P}_{t}}\int^{1}_{0}\ud s_{c} \frac{\ud \trm{P}_{c}[s_{c};\eta_{p}]}{\ud s_{c}} \int^{s_{c}}_{0} \frac{\ud s_{b}}{s_{c}}
\rho_{b}[s;\eta,\bm{\Gamma}_{c}(s_{c})]\,,
\label{Eq_tri_densmatrix}
\end{align}
with $\rho_{b}[s,\eta,\bm{\Gamma}_{c}(s_{c})]$ calculated from Eq.~(\ref{Eq_densmatrix_plane}) without the $s$-integration and normalizing factor $\ud \trm{P}_{b}/\ud s$. Again, the concurrence $\mC(s_{c})$ and $\mC(s_{c},s_{b})$ can be calculated from Eq.~(\ref{Eq_tri_densmatrix}) by removing the corresponding pre-integrals and normalizing with the factor $\ud \trm{P}_{t}/\ud s_{c}$ and $\ud^{2}\trm{P}_{t}/\ud s_{c}\ud s_{b}$, respectively.

\begin{figure}[t!!!]
\center{\includegraphics[width=0.49\textwidth]{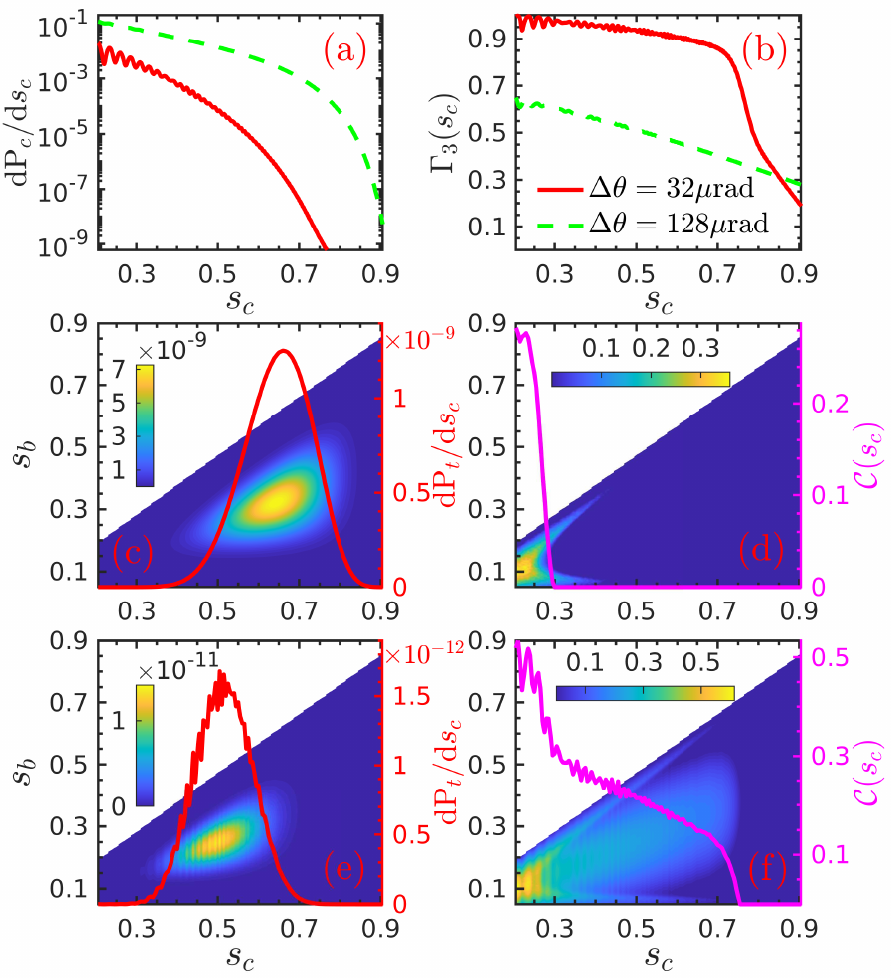}}
\caption{Spin entanglement between the electron-positron pair created in the CP set-up. The both laser pulses have $4$ cycles, an intensity $\xi=2$ and wavelength $0.8~\mu\trm{m}$.
In (a) and (b), the energy spectrum $\ud \trm{P}_{c}/\ud s_{c}$ and polarisation $\Gamma_{3}(s_{c})$ of the Compton generated photon beam are plotted for two cases of finite acceptance angle: $\Delta \theta=32~\mu\trm{rad}$ (red solid  line) and $128~\mu\trm{rad}$ (green solid line).
In (c) and (d), the double lightfront spectrum $\ud^{2}\trm{P}_{t}/\ud s_{c}\ud s_{b}$ and concurrence $\mC(s_{c},s_{b})$ of the electron-positron pairs created by the photon beam with angular acceptance $\Delta \theta=128~\mu\trm{rad}$ are plotted.
Dependence of the probability $\ud\trm{P}_{t}/\ud s_{c}$ and concurrence $\mC(s_{c})$ on the lightfront momentum of the photon is also presented in (c) and (d) on the right-hand vertical axis respectively.
In (e) and (f), the same quantities are plotted as in (c) and (d) but with the photon beam acceptance decreased to $\Delta \theta=32~\mu\trm{rad}$.
}
\label{Fig5_TTri_cir}
\end{figure}

\begin{figure}[t!!!]
\center{\includegraphics[width=0.49\textwidth]{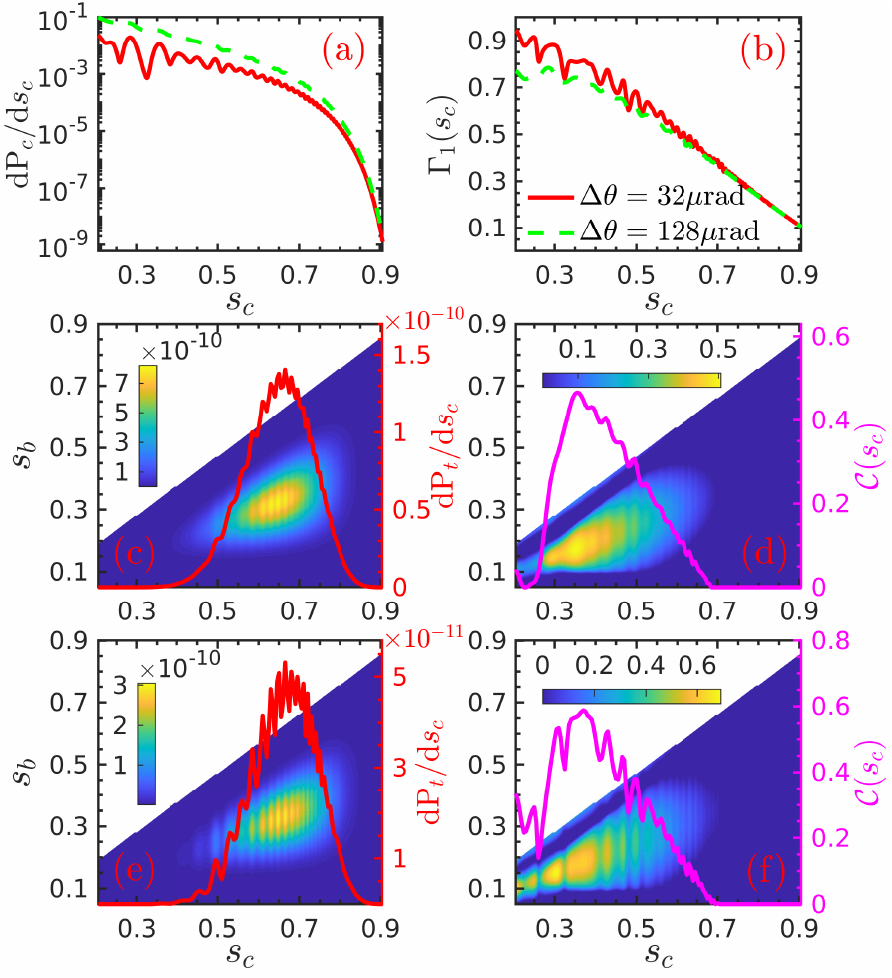}}
\caption{Spin entanglement created with laser and electron beam parameters as in \figref{Fig5_TTri_cir} but now in the LP set-up.
In (a) and (b), the energy spectrum $\ud \trm{P}_{c}/\ud s_{c}$ and polarisation $\Gamma_{1}(s_{c})$ of the Compton photon beam generated in the first laser pulse are given for two cases of finite acceptance angles: $\Delta \theta=32~\mu\trm{rad}$ and $128~\mu\trm{rad}$.
In (c) and (d), the double lightfront spectrum $\ud^{2}\trm{P}_{t}/\ud s_{c}\ud s_{b}$ and concurrence $\mC(s_{c},s_{b})$ of the electron-positron pair created by the photon beam within the angular acceptance $\Delta \theta=128~\mu\trm{rad}$ are plotted.
Dependence of the probability $\ud\trm{P}_{t}/\ud s_{c}$ and concurrence $\mC(s_{c})$ on the lightfront momentum of the photon are also presented in (c) and (d) with the right-hand vertical axis respectively.
In (e) and (f), the same quantities are plotted as in (c) and (d) with the photon beam acceptance decreased to $\Delta \theta=32~\mu\trm{rad}$.}
\label{Fig6_TTri_lin}
\end{figure}

In the two-stage CP set-up, a $16.5~\trm{GeV}$ electron beam is collided with a $4$-cycle CP laser pulse with intensity parameter $\xi=2$, rotation parameter ($\mathfrak{c}=+1)$ and frequency $1.55\,\trm{eV}$, corresponding to an electron energy parameter $\eta_{p}=0.195$. 
As shown in Figs.~\ref{Fig5_TTri_cir} (a) and (b), in stage I) high-energy and highly polarised photons are obtained via nonlinear Compton scattering.
By narrowing the angular acceptance $\Delta \theta$, the degree of the photon polarisation can be effectively improved at the cost of reducing the accepted photon yield~\cite{tang2020highly}.
The scattered photons are then collided with a second laser pulse with the same parameters as in stage I) but with opposite rotation, $\mathfrak{c}=-1$, to generate the electron-positron pair spectrum seen in Fig.~\ref{Fig5_TTri_cir} (c), plotted as the double light-front spectrum $\ud^{2}\trm{P}_{t}/\ud s_{c}\ud s_{b}$ of the pairs created by the photon beam within the acceptance $\Delta \theta=128~\mu\trm{rad}$. The corresponding concurrence $\mC(s_{c},s_{b})$ is then plotted in Fig.~\ref{Fig5_TTri_cir} (d). Because of the low degree of photon polarisation $\Gamma_{3}<0.7$ in Fig.~\ref{Fig5_TTri_cir} (b), the spin of the created particles is only partially entangled in the low-energy region $s_{c}<0.25$, \emph{i.e.} the photon energy $\eta=s_{c}\eta_{p}<0.05$, because of the perturbative effect with the partial photon polarisation as discussed in Fig.~\ref{Fig3_eta_cir} (b).
For higher photon energies ($s_{c}>0.4$) the concurrence falls to zero, implying the pair's spin states are factorisable as the product of spin states of single particles.
To improve the spin entanglement, one may narrow the angular acceptance for higher photon polarisation, \emph{e.g.}, $\Delta \theta=32~\mu\trm{rad}$ shown in Fig.~\ref{Fig5_TTri_cir} (e).
These highly polarised photons would then create the pairs with much stronger spin entanglement shown in Fig.~\ref{Fig5_TTri_cir} (f), but a considerably reduced total pairs (see Fig.~\ref{Fig5_TTri_cir} (e)).

One may also note that the main part of the created pairs with the probability $\ud\trm{P}_{t}/\ud s_{c}> 10^{-12}$ by the photons within $\Delta \theta=32~\mu\trm{rad}$, see the red solid line in Fig.~\ref{Fig5_TTri_cir} (e) around the photon energy~$s_{c}=0.5$,
could contribute to the measurement of spin entanglement with the concurrence $\mC(s_{c})>0.2$, magenta solid line in Fig.~\ref{Fig5_TTri_cir} (f).
However for the photon within $\Delta \theta=128~\mu\trm{rad}$, the only pairs, which can contribute to the spin-entanglement measurement with $\mC(s_{c})>0.1$, are created by the photons with the energy $s_{c}<0.25$ and probability $\ud\trm{P}_{t}/\ud s_{c}<3.5\times 10^{-14}$, which would thus significantly reduce the measurement efficiency for the pairs' spin entanglement.

The LP set-up is the same as for the CP case, except with linearly polarised photons, and the analysis is shown in Fig.~\ref{Fig6_TTri_lin}. As can be seen, the concurrence can reach as large as $\mC \approx 0.6$.
The photon polarisation $\Gamma(s_c)$, and thus the concurrence, are much less affected by the change of angular acceptance compared to the CP case.
For both the CP and LP cases we see the probability for pair-creation maximal near $s_c\sim0.5\!-\!0.7$, and the concurrence reducing to zero for $s_c$ close to $1$.
However the important difference is in the overlap of regions of large concurrence with regions where pair-creation is more probable.
The concurrence $\mathcal{C}(s_c,s_b)$ here remains sizeable for $s_c \gtrsim 0.5$  for both $\Delta\theta=32$ and $128~\mu$rad, where the overlap with the probability $\ud^{2}\trm{P}_{t}/\ud s_{c}\ud s_{b}$ for pair-creation is also large for both acceptances.
This is in contrast with the CP case, where the overlap between $\mathcal{C}(s_c,s_b)$ and the probability $\ud^{2}\trm{P}_{t}/\ud s_{c}\ud s_{b}$ is highly suppressed for the $\Delta\theta=128~\mu$rad, and comparable to LP case for $\Delta\theta=32~\mu$rad.
Therefore at larger angular acceptance, i.e. smaller pulse separations, the LP scenario appears more favourable for observing concurrence in the NBW pair creation, while at smaller angular acceptance the difference becomes less important. This figure can potentially be increased by optimising beam parameters for the largest concurrence and this could potentially form the subject of future work.

\section{Conclusion}~\label{Sec_5}

We have studied the entanglement of electron-positron pairs produced in the collision between a high-energy photon and a high-intensity laser pulse via the nonlinear Breit-Wheeler process. Several parameter regimes have been considered that could be probed by upcoming experiments, such as LUXE \cite{Abramowicz:2021zja}.
We investigated the concurrence measure of spin-polarisation entanglement in an intensity range spanning the perturbative $\xi \ll 1$ to non-perturbative $\xi\gg 1$, energies ranging from low values $\eta \ll 1$ to the high energy range of harmonic pair creation {$\eta > 2(1+\upsilon\xi^{2})$}, for a range of photon polarisations in a circularly-polarised or linearly-polarised background.
After considering mono-energetic photons, we presented results for an experimental scenario featuring two-stages: I) where photons are produced via nonlinear Compton scattering by a first laser pulse followed by II) where the photons collide with a second laser pulse to create pairs.

It was found that the concurrence, $\mC$, depends in a complicated way on laser intensity, photon energy and polarisation.
Of particular interest was the finding that increasing the intensity of the plane wave background can sometimes lead to an \emph{increase} in the concurrence. For example, in a linearly-polarised background, the concurrence can reach values as high as $\mC \approx 0.9$ in the \emph{intermediate} intensity regime i.e. at laser parameters that are available in experiment today. In the two-stage scenario which used a Compton distribution of photons, the concurrence could be maintained at a level as high as $\mC\approx 0.6$ using angular cuts in the photon spectrum that could be realised in experiment e.g. by using collimators, and this figure can potentially be optimised upwards.

Local approximations (the LCFA and LMA) were derived for the spin-polarisation density matrix of the produced pair and benchmarked with values calculated directly from QED.
Although there was in general good agreement in the intensity regimes in which these approximations are expected to be valid, it was again demonstrated (as has been noted in e.g. \cite{BenPRA2020} and \cite{King:2023eeo}) that the LCFA failed to fully capture the physics in a circularly-polarised background. The results can be used to determine the experimental parameters for which these local approximations can be employed in numerical simulations of particles scattering in intense EM fields. In turn, this may help us understand what effect entanglement has on more complicated strong-field processes, such as the development of QED cascades.

Recent developments in particle physics phenomenology have led to the measurement of entanglement at high energies; our work opens the door to probing entanglement in electromagnetic backgrounds at high intensities.
The measurement of entanglement at high-intensity would provide a new test of quantum field theory, potentially shedding light on new physics effects.
One application of our results is in ghost imaging~\cite{PRD056015} in which an idler particle is used to gain information about a target without directly interacting with it, by being entangled with another particle that probes the target.

The approach we have outlined in the current paper can be further used to study other strong-field QED processes, in particular higher-order processes that could lead to entanglement of larger numbers of particles. It would also be important to consider more real-world effects that would be present in any experimental realisation, such as the method chosen for spin measurement and the influence of laser focussing.

\section{Acknowledgments}
Suo Tang acknowledge supported from the Shandong Provincial Natural Science Foundation, Grants No. ZR2021QA088.

\onecolumngrid
\appendix

\section{Different Spin basis}\label{App_spinbasis}

\subsection{Light-front spin quantization axis}
One can define a covariant spin basis vector: lightfront helicity,
\begin{align}
S^{\mu}_{p}&=\frac{p^{\mu}}{m}-\frac{m}{k\cdot p}k^{\mu}\,,
\label{Eq_lightfront_helicity}
\end{align}
in the lab reference, corresponding to the spin quantization axis
\begin{align}
\bm{n}=\frac{p^{+}+m}{p^{+}(p^{0}+m)}(p^{1},p^{2},p^{3}+m\frac{p^{0}+m}{p^{+}+m})\,,
\end{align}
antiparallel to the laser propagating direction in the particle rest frame, and
where $p^{+}=p^{0}+p^{3}=k\cdot p/k^{0}$. For a positron, the momentum $p^{\mu}$ should be replaced with $q^{\mu}$.

In terms of the Chiral (Weyl) representation~\cite{nagashima2011elementary,schwartz2014quantum,peskin2018introduction}:
\begin{align}
\gamma^{0}=\begin{bmatrix}
    0 & \mathds{1}_{2\times2}  \\
    \mathds{1}_{2\times2}  & 0
\end{bmatrix}\,,~~
\gamma^{1}=\begin{bmatrix}
    0 & \hat{\sigma}_1 \\
   -\hat{\sigma}_1 & 0
\end{bmatrix}\,,~~
\gamma^{2}=\begin{bmatrix}
    0 & \hat{\sigma}_2 \\
   -\hat{\sigma}_2 & 0
\end{bmatrix}\,,~~
\gamma^{3}=\begin{bmatrix}
    0 & \hat{\sigma}_3 \\
   -\hat{\sigma}_3 & 0
\end{bmatrix}\,,
\end{align}
where $\hat{\sigma}_{1,2,3}$ are the Pauli matrices, the bispinors of electron ($u_{p,\sigma}$) and positron ($v_{q,\sigma}$) can be given as:
\begin{equation}
\begin{aligned}
 u_{p,+1}&=\frac{1}{\sqrt{2mp^{+}}}\begin{bmatrix}
   m\\
   0\\
   p^{+}\\
   p^{1}+ip^{2}
\end{bmatrix}\,,~~
u_{p,-1}=\frac{1}{\sqrt{2m p^{+}}}\begin{bmatrix}
   ip^{2}-p^{1}\\
   p^{+}\\
   0\\
   m
\end{bmatrix}\,,\\
v_{q,+1}&=\frac{1}{\sqrt{2 m q^{+}}}\begin{bmatrix}
    q^{1}-iq^{2} \\
   -q^{+} \\
    0 \\
    m
\end{bmatrix}\,,~~
v_{q,-1}=\frac{1}{\sqrt{2 m q^{+}}}\begin{bmatrix}
 m \\
 0 \\
 -q^{+}\\
 -(q^{1}+iq^{2})
\end{bmatrix}\,.
\end{aligned}
\label{Eq_spin_basis_lightfront}
\end{equation}

\subsection{Spin quantization axis along $z$-direction}
One can also define the spin quantization axis in the particle's rest frame as $\bm{n}=(0,0,1)$ along the $z$-direction.
Making use of the standard gamma-matrix
\begin{align}
\gamma^{0}&=\begin{bmatrix}
    \mathds{1}_{2\times2} & 0  \\
    0 &-\mathds{1}_{2\times2}
\end{bmatrix}\,,~~
\gamma^{1}=\begin{bmatrix}
    0 & \hat{\sigma}_1 \\
   -\hat{\sigma}_1 & 0
\end{bmatrix}\,,~~
\gamma^{2}=\begin{bmatrix}
    0 & \hat{\sigma}_2 \\
   -\hat{\sigma}_2 & 0
\end{bmatrix}\,,~~
\gamma^{3}=\begin{bmatrix}
    0 & \hat{\sigma}_3 \\
   -\hat{\sigma}_3 & 0
\end{bmatrix}\,,
\end{align}
the explicit expressions of electron ($u_{p,\sigma}$) and positron ($v_{q,\sigma}$)bispinors can be given as
\begin{equation}
\begin{aligned}
 u_{p,+1}&=\sqrt{\frac{p^{0}+m_{0}}{2m_{0}}}
\begin{bmatrix}
 1 \\
 0 \\
\frac{p_{z}}{p^{0}+m_{0}} \\
\frac{p_{x} + i p_{y}}{p^{0}+m_{0}}
\end{bmatrix}\,,~~
u_{p,-1}=\sqrt{\frac{p^{0}+m_{0}}{2m_{0}}}
\begin{bmatrix}
 0 \\
 1 \\
 \frac{p_{x} - i p_{y}}{p^{0}+m_{0}} \\
 \frac{-p_{z}}{p^{0}+m_{0}}
\end{bmatrix}\,,\\
v_{p,+1}&=\sqrt{\frac{p^{0}+m_{0}}{2m_0}}
\begin{bmatrix}
\frac{p_{x} - i p_{y}}{p^{0}+m_{0}}\\
\frac{-p_{z}}{p^{0}+m_{0}}\\
    0 \\
    1
\end{bmatrix}\,,~~
v_{p,-1}=\sqrt{\frac{p^{0}+m_{0}}{2m_0}}
\begin{bmatrix}
\frac{-p_{z}}{p^{0}+m_{0}} \\
-\frac{p_{x} + i p_{y}}{p^{0}+m_{0}} \\
    -1 \\
    0
\end{bmatrix}\,.
\end{aligned}
\label{Eq_spin_basis_zdir}
\end{equation}

\subsection{Numerical comparison}
In plane wave backgrounds, the scattering matrix element can be written out explicitly,
\begin{align}
S_{p\sigma;q\varsigma;\ell\lambda}&=\langle e^{\LCm};p,\sigma|\otimes\langle e^{\LCp};q,\varsigma|\hat{S}|\gamma;\ell,\varepsilon\rangle\nonumber\\
                                  &=-ie\int \ud^4x \overline{\varPsi}_{p,\sigma}(x)\slashed{A}_{\trm{ph}} \varPsi^{+}_{q,\varsigma}(x)\,,
\end{align}
where $\varPsi^{\LCp}_{q,\varsigma}$($\varPsi_{p,\sigma}$) is the Volkov wave function of the produced positron (electron) with the momentum $q^{\mu}$ ($p^{\mu}$) and the spin $\varsigma/2$ ($\sigma/2$)~\cite{wolkow1935klasse}:
 \begin{align}
\varPsi_{p,\sigma}(x)&=\sqrt{\frac{m}{Vp^{0}}}\left(1-\frac{\slashed{k}\slashed{a}}{2k\cdot p}\right)u_{p,\sigma}e^{-ip\cdot x+i\int^{\phi}\ud\phi'
\frac{2p\cdot a+a^2}{2k\cdot p} },\nonumber\\
\varPsi^{\LCp}_{q,\varsigma}(x)&=\sqrt{\frac{m}{Vq^{0}}}\left(1+\frac{\slashed{k}\slashed{a}}{2k\cdot q}\right)v_{q,\varsigma}e^{iq\cdot x+i\int^{\phi}\ud\phi'\frac{2q\cdot a-a^2}{2k\cdot q} }.\nonumber
\end{align}
and $A_{\textrm{ph}}$ is the photon field:
\begin{align}
A^{\mu}_{\textrm{ph}}=\sqrt{\frac{2\pi}{\ell^{0} V}}\varepsilon^{\mu}_{\lambda}e^{-i\ell\cdot x}\,,
\label{Eq_NBW_photon0}
\end{align}
After simple derivation, one can get:
\begin{align}
S_{p\sigma;q\varsigma;\ell\lambda}&=\frac{-ie}{k^{0}}\sqrt{\frac{2\pi m^{2}}{V^{3}q^{0}p^{0} \ell^{0}}}(2\pi)^{3}\delta^{\LCperp,\LCp}(p+q-\ell)\nonumber\\
                                  &~\int \ud \phi~\bar{u}_{p,\sigma}M(\varepsilon_{\lambda},\phi)v_{q,\varsigma} ~ e^{i\int^{\phi}d\phi' \frac{\ell\cdot\pi_{q}(\phi')}{k\cdot p}}
\end{align}
where the lightfront $\delta$-function $\delta^{\LCperp,\LCp}(p+q-\ell)$ guarantees the energy-momentum conservation, and
\begin{align}
M(\varepsilon_{\lambda},\phi)&=\slashed{\varepsilon}_{\lambda}+\frac{\slashed{\varepsilon}_{\lambda}\slashed{k}\slashed{a}(\phi)}{2k\cdot q} -\frac{\slashed{a}(\phi)\slashed{k}\slashed{\varepsilon}_{\lambda}}{2k\cdot p}\,.
\label{Eq_Creation_matrix0}
\end{align}
The final spin density matrix and production probability can then be written as
\begin{align}
\rho_{b}
&=\frac{1}{\trm{P}_{b}}\sum_{\lambda\lambda'}\rho_{\gamma,\sscript\lambda\lambda'}\frac{\alpha}{(2\pi\eta)^2}\int \frac{\ud s}{ts} \int\ud^{2} \bm{r} \iint \ud \phi_{1} \ud \phi_{2}\nonumber\\
&~~~~~~~e^{i\int_{\phi_{2}}^{\phi_{1}}d\phi'\frac{\ell\cdot\pi_{q}(\phi')}{m^{2}\eta t}}~ \textrm{T}_{\sigma\varsigma\lambda;\sigma'\varsigma'\lambda'}(\phi_1,\phi_{2})\,,
\label{Eq_finialdens_matrix}
\end{align}
and
\begin{align}
{\trm{P}_{b}}
&= \sum_{\sigma,\varsigma}\sum_{\lambda\lambda'}\rho_{\gamma,\sscript\lambda\lambda'}\frac{\alpha}{(2\pi\eta)^2}\int \frac{\ud s}{ts} \int\ud^{2} \bm{r} \iint \ud \phi_{1} \ud \phi_{2}\nonumber\\
&~~~~~~~e^{i\int_{\phi_{2}}^{\phi_{1}}d\phi'\frac{\ell\cdot\pi_{q}(\phi')}{m^{2}\eta t}}~ \textrm{T}_{\sigma\varsigma\lambda;\sigma\varsigma\lambda'}(\phi_1,\phi_{2})\,,
\label{Eq_prob_App}
\end{align}
where $\textrm{T}_{\sigma\varsigma\lambda;\sigma'\varsigma'\lambda'}=\bar{u}_{p,\sigma}M(\varepsilon_{\lambda},\phi_{1})v_{q,\varsigma}
 \bar{v}_{q,\varsigma'}\overline{M(\varepsilon_{\lambda'},\phi_{2})} u_{p,\sigma'}$
and $\overline{M(\varepsilon_{\lambda'},\phi)}=\gamma^{0}M^{\dagger}(\varepsilon_{\lambda'},\phi)\gamma^{0}$.

\begin{figure}[t!!!]
\center{\includegraphics[width=0.46\textwidth]{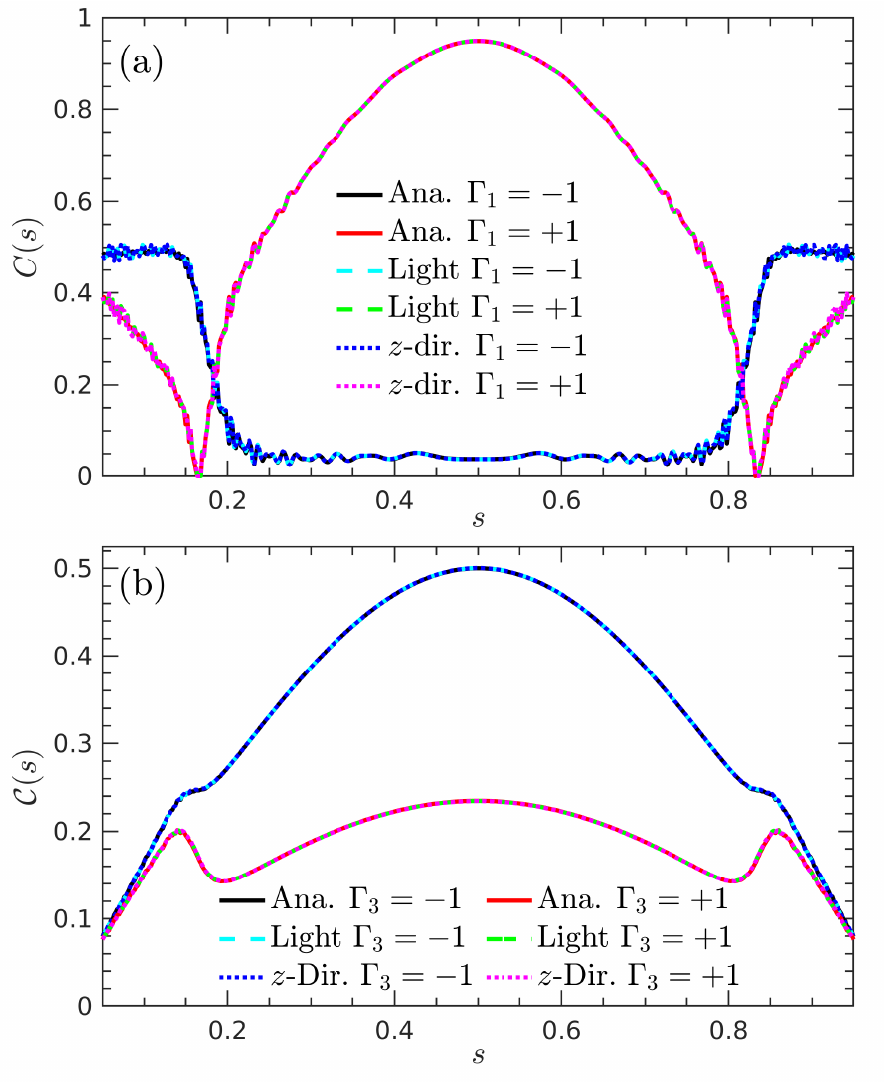}}
\caption{Comparison between the concurrence $\mC(s)$ calculated with different spin bases in the (a) linearly and (b) circularly polarised laser fields for the high-energy photon with $\eta=0.2$.
In (a), the photon is linearly polarised with $\Gamma_{1}=\pm1$ and circularly polarised with $\Gamma_{3}=\pm1$ in (b).
The solid lines denote the results from the analytical expressions in~(\ref{Eq_unpolar_matrix}) and~(\ref{Eq_polar_matrix}), and the dashed and dotted lines show, respectively, the direct calculations with the light-front bispinors and those with the spin axis along the $z$-direction.
The laser pulse has $N=4$ cycles and the intensity $\xi=1$.}
\label{Fig_Entangle_diffspin}
\end{figure}

In the main text, the analytical expressions~(\ref{Eq_unpolar_matrix}) and~(\ref{Eq_polar_matrix}) of the final density matrix is given based on the light-front spin basis.
The energy and transverse-momentum distribution of the final density matrix $\rho_{f}(s)$ and $\rho_{f}(r_{x},r_{y})$ can also be acquired by removing the corresponding pre-integrals in~(\ref{Eq_finialdens_matrix}) and normalizing respectively with $\ud \trm{P}/\ud s$ and $\ud^{2}\trm{P}/\ud r_{x}\ud r_{y}$, and could be used to measure the entanglement between the produced particles with different energy and transverse momentum by calculating the concurrence $\mC(s)$ and $\mC(r_{x},r_{y})$.

With the above bispinors~(\ref{Eq_spin_basis_lightfront}) and~(\ref{Eq_spin_basis_zdir}), one can calculate the final density matrix~(\ref{Eq_finialdens_matrix}) directly and then acquire the concurrence of the density matrix.
As shown in Fig.~\ref{Fig_Entangle_diffspin}, the energy spectra of the concurrence $\mC(s)$ from the analytical results~(\ref{Eq_polar_matrix}) match exactly with the numerical results calculated directly with the light-front spin basis, and also match very well with those calculated with the bispinors corresponding to the spin quantization axis along $z$-direction.
As one can see, the energy spectra of the concurrence $\mC(s)$ is symmetric at $s=0.5$.
The sharp changes around $s=0.15$ and $s=0.85$ are because of the finite pulse effect.

\section{Compton photon source}\label{App_NLC}
In an electron-laser collision, the differential probability of emitting a photon in the polarisation state $\varepsilon_{\lambda}$ with momentum $\ell$ via the nonlinear Compton process, can be written as~\cite{seipt2017volkov,tang2020highly}
\begin{align}
\frac{\ud\trm{P}_{c,\lambda}}{\ud s_{c}} =& \frac{\alpha}{(2\pi\eta_{p})^{2}}\frac{s_{c}}{1-s_{c}}\iint \frac{\ud^{2} \bm{\ell}^{\LCperp}}{m^{2}s^{2}_{c}}\iint \ud\phi_{1}\,\ud\phi_{2}\nonumber\\
&~\trm{T}_{\lambda}(\phi_{1},\phi_{2})~e^{i\int^{\phi_{1}}_{\phi_{2}}\ud\phi\frac{k\cdot \pi_{p}(\phi)}{m^2(1-s_{c})\eta_{p}}},\label{eqn:sfi1}
\end{align}
where $\eta_{p} = k\cdot p/m^{2}$, $s_{c}=k\cdot \ell /k\cdot p$ is the lightfront momentum fraction of the scattered photon, $\pi_{p}=p + a - k(2\,p\cdot a + a^{2})/k\cdot p$ is the instantaneous momentum of the electron, and $\trm{T}_{\lambda}$ is a polarisation-dependent integrand.
The parameter $\bm{\ell}^{\LCperp}/(s_{c}m)$ is the normalised photon momentum in the plane perpendicular to the laser propagation direction and relates directly to the scattering angles of the photon. If 
$\theta$ is the angle to the positive $z$ axis,  and $\psi$ the transverse polar angle, then  $\bm{\ell}^{\LCperp}/(s_{c}m) =m\eta_{p}\tan(\theta/2)/k^0(\cos\psi,\sin\psi)$. 
For a specified acceptance $\Delta \theta$, the integral over the transverse momentum in Eq.~(\ref{eqn:sfi1}) would be confined in the region corresponding to the polar angle $0\leq\theta\leq\Delta \theta/2$.

The photon polarisation states are chosen to be the eigenstates of the polarisation operator in the given laser background~\cite{baier76}, as
\[\varepsilon_{1} = \epsilon_{1} - \frac{\ell\cdot \epsilon_{1}}{k\cdot \ell} k\,,~~~
  \varepsilon_{2} = \epsilon_{2} - \frac{\ell\cdot \epsilon_{2}}{k\cdot \ell} k, \]
for a linearly-polarised background, and a circularly polarised background
\[\varepsilon_{\pm} = (\varepsilon_{1}\pm i\varepsilon_{2})/\sqrt{2} \]
where $\epsilon_{1} = (0,1,0,0)$ and $\epsilon_2=(0,0,1,0)$. The integrand $\trm{T}_{\lambda}$ for the photon polarisation $\varepsilon_{\lambda}$ can then be given as:
\begin{subequations}
\begin{align}
\trm{T}_{1/2}(\phi_{1},\phi_{2}) &= \frac{s^2_{c}(\Delta a)^{2}}{8(1-s_{c})} +w_{c,1/2}(\phi_{1})\cdot w_{c,1/2}(\phi_{2})\,,\nonumber\\
\trm{T}_{\pm}(\phi_{1},\phi_{2}) &= \frac{s^2_{c}(\Delta a)^{2}}{8(1-s_{c})} +\frac{1}{2}\bm{w}_{c}(\phi_{1})\cdot\bm{w}_{c}(\phi_{2})\nonumber\\
              &~~~~~~~~~~~~~~~~ \pm i f_{s}\bm{w}_{c}(\phi_{1})\times \bm{w}_{c}(\phi_{2}),\nonumber
\end{align}
\end{subequations}
$\!\!$with $\bm{w}(\phi) =\bm{\ell}^{\LCperp}/(s_{c}m)-\bm{p}^{\LCperp}/m - \bm{a}^{\LCperp}(\phi)/m$, $\Delta a=[a(\phi_{1})-a(\phi_{2})]/m$, $f_s=(2 - 2s_{c} + s_{c}^2)/[4(1-s_{c})]$.

The total spectrum of the emitted photon can be written as $\ud\trm{P}_{c}/\ud s_{c} = \ud\trm{P}_{c,1}/\ud s_{c} + \ud\trm{P}_{c,2}/\ud s_{c}$
in a linearly polarised background, and $\ud\trm{P}_{c}/\ud s_{c} = \ud\trm{P}_{c,\LCp}/\ud s_{c} + \ud\trm{P}_{c,\LCm}/\ud s_{c}$ in  a circularly polarised background.
The polarisation degree of the emitted photons is define in the linear case as
\begin{align}
\Gamma_{1}(s_{c}) = \frac{\ud\trm{P}_{1}/\ud s_{c} - \ud\trm{P}_{2}/\ud s_{c}}{\ud\trm{P}_{1}/\ud s_{c} + \ud\trm{P}_{2}/\ud s_{c}}
\end{align}
and
\begin{align}
\Gamma_{3}(s_{c}) = \frac{\ud\trm{P}_{\LCp}/\ud s_{c} - \ud\trm{P}_{\LCm}/\ud s_{c}}{\ud\trm{P}_{\LCp}/\ud s_{c} + \ud\trm{P}_{\LCm}/\ud s_{c}}
\end{align}
in the circular case, same as the definition in Eq.~(\ref{Eq_phpolar}) in the main text.

\twocolumngrid

\bibliographystyle{apsrev}
\bibliography{NBW_Entanglment}
\end{document}